\documentclass[a4paper,12pt]{article}
\usepackage[utf8]{inputenc}
\usepackage{cancel}
\usepackage{ulem}
\usepackage{breqn}
\usepackage{amsfonts}
\usepackage{amssymb}
\usepackage{graphicx}
\usepackage{amsmath}
\usepackage{enumerate}
\usepackage{mathtools}
\usepackage{subfig}
\usepackage{color}
\usepackage{tikz}
\usepackage{float}
\usepackage{here}
\usepackage{cite}
\usepackage{mathrsfs}
\usepackage{float,epsfig}
\usepackage{dcolumn}
\usepackage{lscape}
\usepackage{subfig}
\usepackage{graphicx}
\usepackage{bm}
\usepackage{amsmath,amssymb,amsthm}
\usepackage[colorlinks=true,linkcolor=blue,citecolor=red]{hyperref}
\usepackage{multirow}
\usepackage{amsmath}
\usetikzlibrary{calc}
\newcommand{\be}{\begin{equation}}
\newcommand{\ee}{\end{equation}}
\newcommand{\bea}{\setlength\arraycolsep{2pt} \begin{eqnarray}}
\newcommand{\eea}{\end{eqnarray}}

\def\0{{\sst{(0)}}}
\def\1{{\sst{(1)}}}
\def\2{{\sst{(2)}}}
\def\3{{\sst{(3)}}}
\def\4{{\sst{(4)}}}
\def\5{{\sst{(5)}}}
\def\6{{\sst{(6)}}}
\def\7{{\sst{(7)}}}
\def\8{{\sst{(8)}}}
\def\sst#1{{\scriptscriptstyle #1}}

\makeatletter \@addtoreset{equation}{section}

\begin{document}

\title{{\normalsize \textbf{\Large  On RN-AdS Black Holes  with a Cloud of Strings
and  Quintessence  in Noncommuative Geometry}}}
\author{ \small   Saad Eddine Baddis$^{1}$,    Adil  Belhaj$^{1}$, Hajar Belmahi$^{1}$,  Wijdane El Hadri$^{2,3}$, and     Maryem  Jemri$^1$\footnote{maryem.jemri@um5r.ac.ma}\thanks{Authors are listed  in alphabetical order.} \hspace*{-8pt} \\
{\small $^1$ESMaR, Faculty of Science, Mohammed V University in Rabat, Rabat, Morocco  }\\
{\small $^2$ENSIASD, Taroudant,  Ibnou Zohr University, Agadir, Morocco  }\\
{\small $^3$LPTHE, Faculty of Science, Ibnou Zohr University, Agadir, Morocco }
 }
\maketitle

\begin{abstract}
Motivated by certain recent  works, we study thermodynamic and optical properties  of  the Reissner-Nordström-AdS black holes in a noncommutative  spacetime with a string cloud and  quintessence dark  fields.      After analyzing the  global and the  local stabilities, we examine  the criticality and  the Joule-Thomson expansion behaviors in such noncommutative   backgrounds.        Concretely,  we find that the   critical universal numbers $X_N$ can be expressed as $X_0$ + $NX$, where $N$ is the quintessence parameter and $X_0$ is the critical universal number of  the Reissner-Nordström-AdS black holes in  a NC spacetime without additional external fields.  Furthermore, we find that Van der Waals behaviors can be recovered by strictly constraining the charge in terms of external parameters. To conclude this work, we compute  and examine  the  deflection angle of lights   in such  a  modified spacetime geometry.    \\
{\noindent} 
\end{abstract}
\newpage
\tableofcontents
\newpage
\section{Introduction}
 Noncommutative  geometry (NC) has been extensively investigated  in terms of D-branes interpreted as solitonic solutions in string theory \cite{1}. More precisely,  it has appeared in the study of quantum behaviors of D-brane objects  coupled to certain fields of closed string spectrums including the graviton $g_{\mu\nu}$ and the antisymmetric $B_{\mu\nu}$, usually called B-field.   The presence of such fields on the D-brane world-volume  generates non-trivial  commutation relations going beyond the ones appearing in the ordinary quantum mechanics \cite{001,up12,up15,up11,up13,up14,up16}.   In string theory and related topics,  this scenario has   provided a NC spacetime  deformed by an antisymmetric  tensor linked to the inverse of the B-field belonging to the NS-NS sector. These NC  behaviors    have been largely studied in connection with several 
 subjects including   quantum field and gauge theories \cite{2,up9,212,2121,adilsaidi}.  Moreover, they have been introduced  in the study of Calabi-Yau manifolds where the  NC  parameters could be exploited to remove  singular aspects \cite{3,adil1,adil2,adil3}.

Recently,   the black holes on NC spaces   have been considered as relevant topics in connection with recent developments in optical and thermodynamical behaviors  of certain gravity models \cite{4,45,5,6,66,77}.
It has been shown that these geometries introduce quantum corrections to black hole 
behaviors, including  optics  and thermodynamics. Concerning 
thermodynamics, many quantities have been computed and examined for several 
black holes in different gravity models. For Schwarzschild black holes in  the 
NC spacetime, for instance, the point-like mass is replaced by a 
Gaussian distribution with a minimal length scale, smoothing the geometry 
and modifies the thermodynamical behaviors \cite{100,1001,up17}. It has been observed that the Hawking 
temperature exhibits similarities with that of an ordinary charged black hole 
possessing two horizons \cite{7,8}. A similar 
discussion has been established for the  Reissner--Nordström black holes, where the 
noncommutative parameter plays a role analogous to the electric charge\cite{A7}. 

The optical properties of the  black holes in the NC spacetime have 
been  also investigated, showing that the noncommutative  parameter  alters the propagation 
of lights near the black hole. In models inspired by  the NC  geometry, 
the size of the black hole shadow generally decreases and its shape is distorted 
compared to the commutative case~\cite{9}. Research  activities  
noncommutative Schwarzschild black holes have  shown that stronger noncommutative 
effects lead to a reduction in the shadow radius~\cite{10,up10}. 
Comparable results have been elaborated in more sophisticated frameworks, such 
as noncommutative Einstein--Gauss--Bonnet black holes, where both the shadow 
size and the light deflection angles are markedly influenced by the 
NC  parameter \cite{ref3}.

 More recently,  the Reissner--Nordström--AdS  black holes in a NC  spacetime with  Lorentzian-smeared distributions have been investigated where a particular emphasis  has been  put on   thermodynamic  properties\cite{WM}.  By computing the  critical  thermodynamic quantities,  it has been shown that such black holes  exhibit   certain similarities with   Van
der Waals fuild systems.  Moreover, it has been revealed that  the Joule–Thomson expansion computations    have provided   perfect universalities   appearing in  the   ordinary charged AdS black holes.

 In this work, we study   the Reissner--Nordström--AdS  black holes in  a NC  spacetime with a cloud of strings and quintessence   dark sector  fields. 
Using  thermodynamical and optical tools, we show how  such black holes behave in  the presence  of such external   field sources.    
After discussions on  the global and  the local  stabilities,  we   investigate  the    criticality  and  the Joule–Thomson expansion   behaviors   in  a NC space  with a   cloud of strings and quintessence   fields.       More precisely,  we   find that the critical  universal numbers  $X_N$ can be expressed as  $X_0$ + $NX$ where $N$ is the  quintessence  parameter and where  $X_0$ is  the critical  universal of  Reissner--Nordström--AdS  black holes  only  in  a NC  spacetime without extra external fields.  Moreover, we reveal that the Van del Waals behaviors  can be recovered by imposing a strict   constraint on the charge in terms of  the external parameters.  To end this work, we approach  the associated optical behaviors by computing and analyzing    the deflection angle of lights in such backgrounds.

The organization of this work is as follows.  In section 2, we   give a concise discussion on  the Reissner--Nordström--AdS  black holes in a  NC  spacetime with  a cloud of strings and quintessence   dark energy fields.   In section 3, we  examine the  global and the  local stability behaviors.  In section 4, we   investigate
    the criticality and   the Joule-Thomson expansion effects.  In section 5,  we   discuss  the optical properties via the deflection angle variation.
   The last section is devoted to concluding remarks.

\section{Noncommutative  quintessential RN--AdS black holes with a cloud of strings}
\begin{figure}[h!]
    \centering
    \begin{tabular}{cc}
       \includegraphics[width=7.5cm,height=7.5cm]{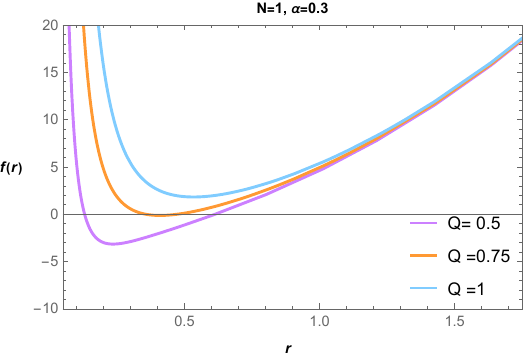} & 
         \includegraphics[width=7.5cm,height=7.5cm]{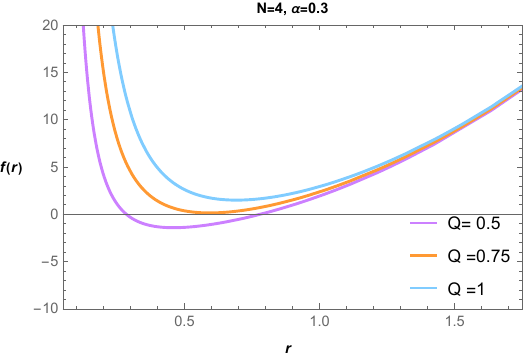} 
    \end{tabular}
     \caption{Effect of the charge parameter $Q$ and  $N$ on the metric function $f(r)$ for $M=1$, $w=-2/3$ and $\Lambda=-20$.}
    \label{Fig3.1}
\end{figure}

Recently, NC spaces have attracted considerable interest in relation to black holes \cite{4}.  Such spaces appear  naturally   in the study of  D-brane objects  in the presence of  the  NS-NS  antisymmetric B-field  of string theory \cite{45}.    Putting  $\hbar=1$, the  coordinates of these  NC spaces are considered  as operators satisfying   the following  commutation relations
\begin{equation}
[x^\mu , x^\nu] = i\,\theta^{\mu\nu},
\end{equation}
where  $\theta^{\mu\nu}$  is    a constant antisymmetric tensor. In large   field approximations,  this tensor has been linked  to the inverse of the  stringy B-field.    Many  field theory models  have been investigated using NC  spaces with the simplified  tensor form 
\begin{equation}
\theta^{\mu\nu}=\Theta  \epsilon^{\mu\nu},
\end{equation}
where  $\epsilon^{\mu\nu}$ is the usual antisymmetric tensor of order 2  and $\Theta$ is  a  NC parameter   having a  length-squared dimension.  In the present work, we reconsider the  study of  charged  black holes on such  NC  spaces  involving only one parameter which  could be related to  a  constant B-field in string theory activities.  This may open gates to study new  black hole solutions in the string theory with D-branes objects and  fields of the  type II spectrum including  the R-R sector.  Roughly,  assuming that the spacetime is static, spheric and symmetric,  the   Reissner--Nordström--AdS  black hole on NC spaces  could be described by the following metric line element
\begin{equation}
 ds^2 = -f(r) dt^2 + \frac{dr^2}{f(r)} + r^2 d\theta^2+  r^2 \sin^2 \theta d\phi^2.
\end{equation}
It turns  out that $f(r)$     is a  relevant radial  function which  can be obtained by solving the Einstein equations with a cosmological constant $\Lambda$
\begin{equation}
 G_{\mu\nu} + \Lambda g_{\mu\nu} = 8\pi T_{\mu\nu},
\end{equation}
where  $ G_{\mu\nu}$ is the Einstein  tensor.    $T_{\mu\nu}$  is the energy-momentum tensor depending  on the studied  black holes.   It has been observed that  the radial function form usually  depends on 
the black hole moduli space ${\cal M}$.  The latter has been shown to be split as follows 
\begin{equation}
{\cal M}= {\cal M}_{int}\times {\cal M}_{ext}.
\end{equation}
Ignoring the rotation parameter, the first factor  called internal moduli space  characterized by  the ordinary parameters 
\begin{equation}
{\cal M}_{int}=\{ M, Q, \Lambda \},
\end{equation}
where   $M$ and  $Q$   are the mass and the charge,  respectively.  The second  factor  is called   external   moduli space involving parameters going beyond the ones  defining $ {\cal M}_{int}$. Such  a model  space concerns  geometrical  and   physical modifications of the spacetime in which black holes live.   It has been  suggested  that   external contributions  have been  motivated by certain  theories including the modified gravity supported by string theory and related topics.  Such contributions have provided many explicit  forms  for the metric function $f(r)$. These activities have furnished  certain predictions  matching with the empirical findings  of EHT collaborations\cite{11,12,13}. Here,  however,  we consider  a specific external moduli space by combining geometric and  matter modificational contributions.  Concretely,  we  deal with a   RN-AdS black hole  with a clould of strings and  quintessence fields  in the  noncommutative geometry   described by the following  external parameters 
\begin{equation}
{\cal M}_{ext}=\{ a, \alpha,  N,  w\},
\end{equation}
where  $\alpha$ denotes    the  cloud string parameter and $a$ is   a parameter dimension of $[L]$ being linked  to  the noncommutativity  one  $\Theta$ via the relation 
\begin{equation}
 a= \frac{8\sqrt{\Theta}}{\sqrt{\pi}}.
\end{equation}
 $N$  and $w$  are  quintessence field parameters with $-1<w<-1/3$. Inspired by many works including the recent ones reported in  \cite{WM,14}, we   deal with     the following metric function
\begin{equation}
 f(r) = 1 -\alpha- \frac{2M}{r} + \frac{aM}{ r^2} + \frac{Q^2}{r^2} - \frac{aQ^2} {r^3}- \frac{\Lambda r^2}{3}-\frac{N}{r^{3w+1}}.
\end{equation}
Ignoring the external moduli space, this function reduces to 
\begin{equation}
 f(r) = 1 - \frac{2M}{r} + \frac{Q^2}{r^2} -\frac{\Lambda r^2}{3},
\end{equation}
representing the  metric function  of a  RN-AdS  black hole  in  the ordinary  spacetime  \cite{15}.
It has been remarked that the thermodynamical and the optical  properties  are encoded in the metric function depending  on the above internal and  the external parameters\cite{150,151,152, 1520,153,154}.   To approach  such  behaviors certain parameters should be fixed generating  a reduced moduli space.   Fixing the mass and the  cosmological constant, for instance,  the discussion can be elaborated in terms  of  a five dimensional moduli space coordinated by  $(Q, a, \alpha, N, w)$.  Fig.(\ref{Fig3.1}), roughly,  shows  such  variation behaviors.
Fixing  the values of   the internal parameters $(a, \alpha, N, w)$,   there exists a critical charge 
$Q_c$  associated with a double root of  the algebraic equation $f(r)=0$. Such a critical value  generates an  extremal black hole solution.  For $Q>Q_C$, we observe  a naked singularity. For  $Q<Q_C$,  however, one   has a non-extremal black hole.  To investigate the physical behaviors of  the proposed  NC   black holes, we   should consider   regions of  the  reduced moduli space  permitting acceptable  solutions of such an algebraic equation. We will refer to  such solutions as  RN-AdS black holes  with a cloud of strings
and   quintessence  fields  in  the NC  geometry.
 
\section{Thermodynamic stability}
\begin{figure}[h!]
    \centering
\begin{tabular}{cc}
\includegraphics[width=8cm,height=6cm]{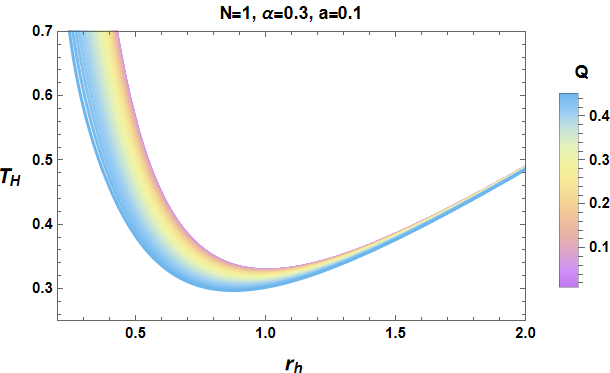} &
\includegraphics[width=8cm,height=6cm]{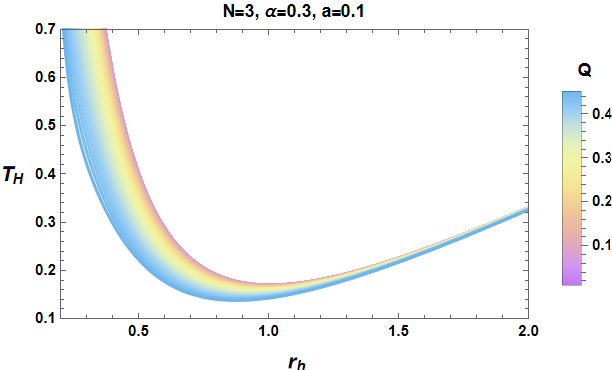}
\end{tabular}
\begin{tabular}{cc}
\includegraphics[width=8cm,height=6cm]{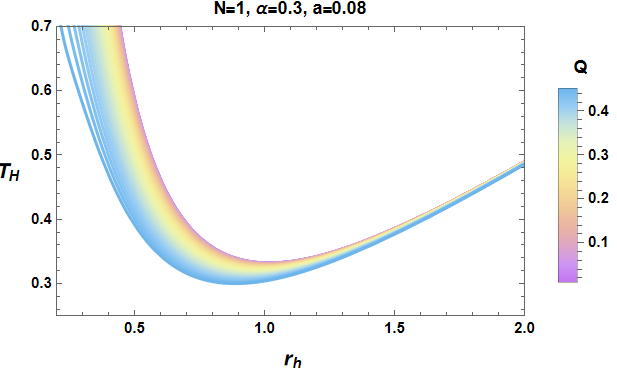} &
\includegraphics[width=8cm,height=6cm]{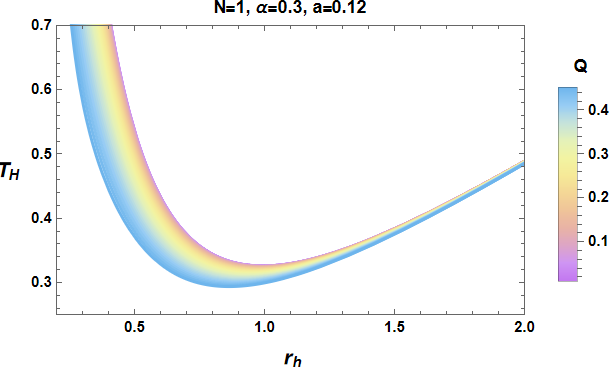}
\end{tabular}
\begin{tabular}{cc}
\includegraphics[width=8cm,height=6cm]{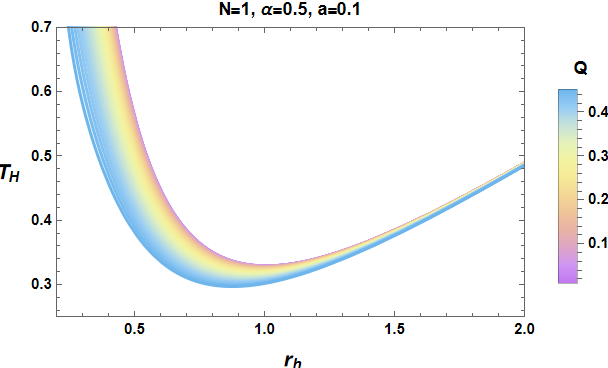} &
\includegraphics[width=8cm,height=6cm]{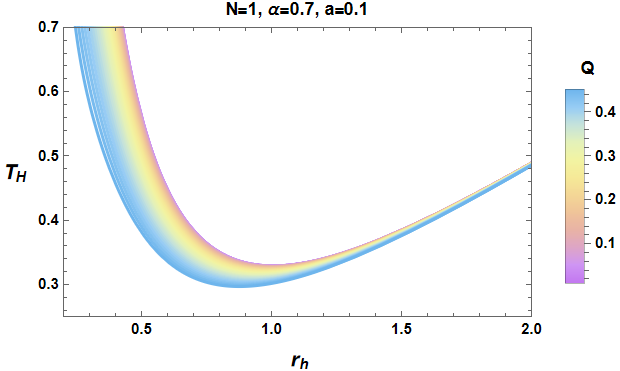}
\end{tabular}
    \caption{Effect the   parameters   $Q$ and $N$ on the Hawking temperature $T$  as a function of $r_h$ by taking  $\Lambda=-5$ and $w=-2/3$.}
    \label{Fig3.2}
\end{figure}

In this section, we delve into the examination  of the local and  the  global   stabilities of such noncommutative  charged black holes in the presence of the above  external parameters.    These  thermodynamical behaviors are encoded in the metric function $f(r)$. The latter   helps to determine the relevant quantities needed to approach such a  stability aspect.  The crucial one is the mass quantity  which can be 
obtained from the constraint $f(r_h)=0$,   where  $r_h$ denotes the horizon radius.   Indeed, the  mass of   the noncommutative RN--AdS black holes  with a cloud of strings
and  quintessence  fields   is shown to be 
\begin{equation}
M =\frac{3 N \,r_h^{2-3 w}+\Lambda  \,r_h^{5}+3\left( \alpha -1\right) r_h^{3} +3 Q^{2} (a-r_h)}{3 \left(a-2 r_h  \right) r_h}.    
\end{equation}
In the absence of $N$ and $\alpha$ external parameters,  we find  the expression obtained in   \cite{WM} being 
\begin{equation}
M =  \frac{\Lambda  \,r_h^{5}+3 Q^{2} a -3 Q^{2} r_h -3 r_h^{3}}{3 r_h \left(a-2 r_h  \right)}.  
\end{equation}
Moreover, taking  $Q=0$,  we recover the    expression 
\begin{equation}
M =\frac{\Lambda  \,r_h^{4}-3 r_h^{2}}{ 3(a-2 r_h) },  
\end{equation}
 representing the mass of  the  Schwarzschild-AdS black hole  in  a  NC spacetime \cite{17}.
The Hawking temperature can be derived using $T_H=\frac{\kappa }{2\pi }$, where $\kappa $
is the surface gravity defined by $\kappa =\frac{1}{2}\frac{\partial f(r)}{%
\partial r}\bigg\rvert_{r=r_{h}}$.    The computations lead to 
\begin{equation}
T_H = \frac{3 N r_h^{2-3w} \left( 3w(a - 2r_h) - a \right) 
+ 6 r_h^3 (\alpha - 1)(r_h - a) 
+ 2 \Lambda r_h^{5} \left( 3 r_h - 2 a \right) 
+ 3 Q^{2} \left( a^{2} - 4 a r_h + 2 r_h^{2} \right)}
{12 \left( a - 2 r_h \right) r_h^{4} \pi}.\label{xxx}
\end{equation}
In the limits $Q=0$ and $N=\alpha=0$, the  Hawking temperature reduces to
\begin{equation}
T_H=\frac{3 \Lambda  \,r_h^{6}-3 r_h^{4}+3 a \,r_h^{3}-2 \Lambda  a \,r_h^{5}}{6 \left(-2 r_h +a \right) r_h^{4} \pi}
\end{equation}
recovering the result obtained in \cite{17}.  Considering  $a= 0$ and $\Lambda= 0 $,  we   obtain  the  temperature of the usual  Schwarzschild black
hole  expressed by   $T_{H}=\frac{1}{4\pi r_{h}}$ \cite{18}.
To unveil the thermal variation,  we  illustrate  the obtained temperature  in terms of the event horizon radius by considering acceptable regions in the reduced moduli space. 
In Fig.(\ref{Fig3.2}),   we plot the Hawking temperature as a function of the
event horizon radius $r_{h}$ for  selected values in such a 
moduli space.

In certain  parametric ranges, it  has been observed that the Hawking temperature  decreases  to a minimal  value.    An examination reveals  that 
the minimal  value decreases   by augmenting  the electric charge $Q$. A similar
behavior has been remarked for   the  external parameters supporting the  existence of  non-trivial  transitions. 
\newpage
\subsection{Global stability}
\begin{figure}[h!]
    \centering
\begin{tabular}{cc}
\includegraphics[width=8cm,height=6.5cm]{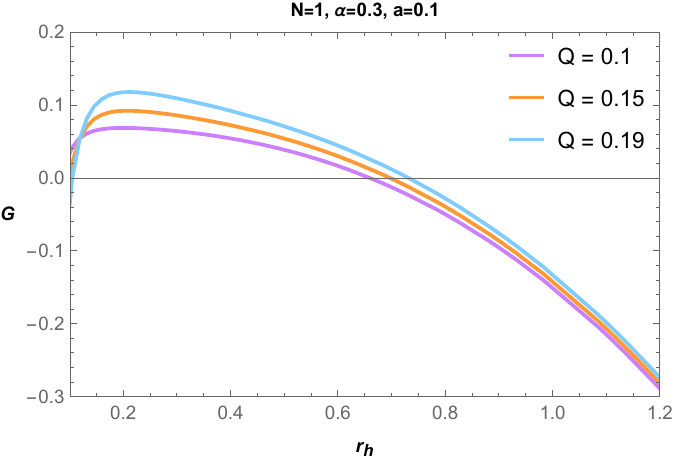} &
\includegraphics[width=8cm,height=6.5cm]{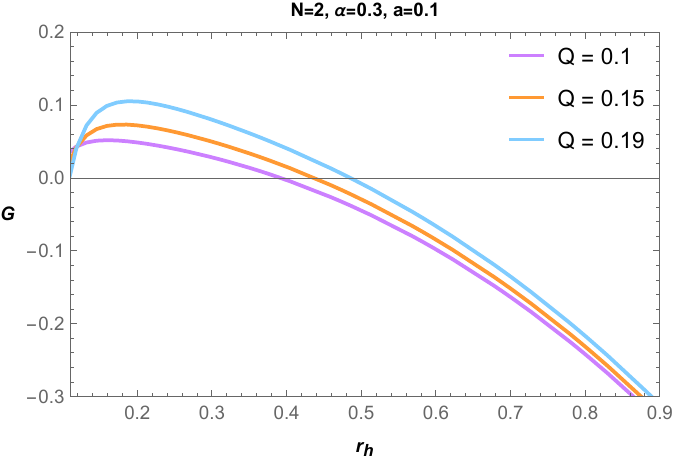}
\end{tabular}
\begin{tabular}{cc}
\includegraphics[width=8cm,height=6.5cm]{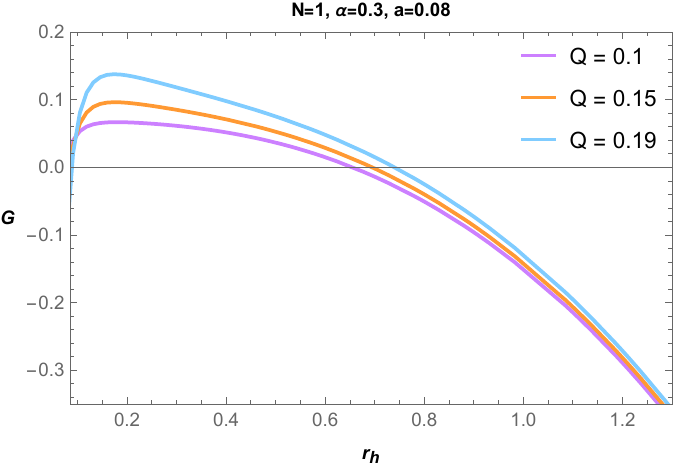} &
\includegraphics[width=8cm,height=6.5cm]{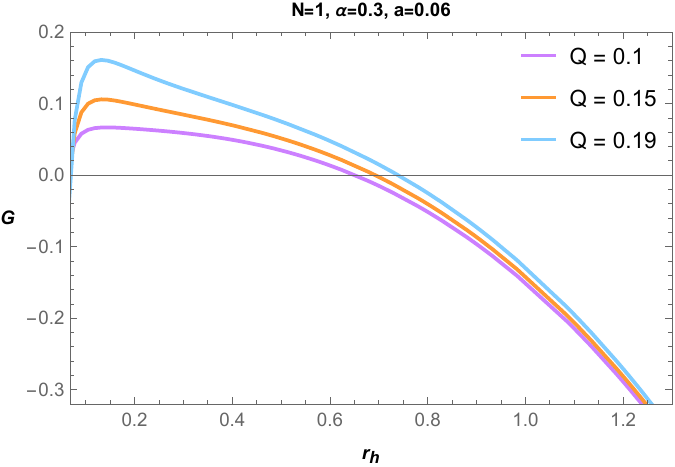}
\end{tabular}
\begin{tabular}{cc}
\includegraphics[width=8cm,height=6.5cm]{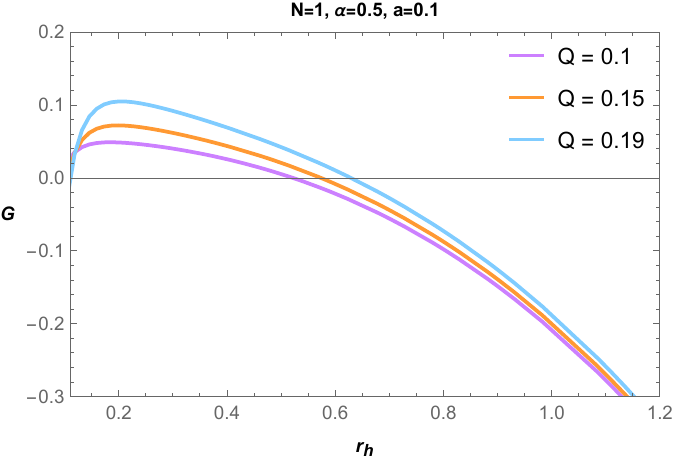} &
\includegraphics[width=8cm,height=6.5cm]{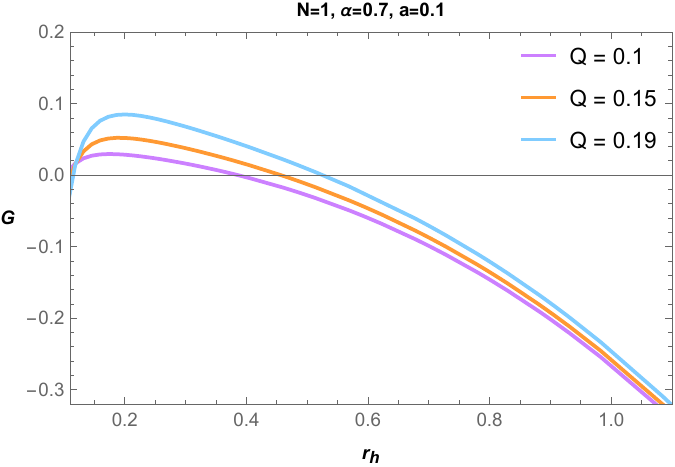}
\end{tabular}
    \caption{Gibbs free energy as a function of $r_h$ with $\Lambda=-0.1$ and $w=-2/3$.}
    \label{7}
\end{figure}

Having discussed the thermal behaviors, we move  to approach the stability behaviors. In particular, we     examine    the global and  the local  stability behaviors   by 
computing the relevant quantities  using appropriate method   used  the  thermodynamic formalism.   First,  we discuss  the  global  stability by approaching   the Gibbs free energy  given by  
\begin{equation}
   G = M - T_H S,
\end{equation}
where $S$ denotes the entropy which  can be obtained  from   the Bekenstein--Hawking area law. Indeed, it can be expressed as 
 \begin{equation}
S =\dfrac{\cal
A}{4} =\pi r_h^2,
\end{equation}
 where  one has  used ${\cal A}=\iint\sqrt{g_{\theta \theta}g_{\phi \phi}} \, d\theta \, d\phi=4\pi r^2_h $ which 
represents  the surface area of the black hole event horizon. 
The computations give
{\footnotesize
\begin{equation}
G =\frac{3 N r_h^{2-3 w} \! \left(\left(1-3 w \right) a +2 \left(3 w +2\right) r_h \right)-2 \Lambda  r_h^{5} \! \left(r_h -2 a \right)+6 \left(\alpha -1\right)  \! \left(r_h +a \right)r_h^{3}-3 Q^{2}(9r_h^2-8ar_h+a^2)}{12 \left(a-2 r_h  \right) r_h^{2}}.
\end{equation}
}
For  $N=\alpha=w=0,$  the Gibbs free energy  takes the form 

\begin{equation}
G = \frac{-2 \Lambda  \,r_h^{5}(r_h-2a)-3 Q^{2}(9r_h^2-8ar_h+a^2)-6 r_h^{3}(r_h+a)}{12r_h^{2} \left(a-2 r_h  \right) }
\end{equation}
being exactly the expression  found in  \cite{WM}.  Taking $Q=0$,  further,  the Gibbs free energy reduces to
\begin{equation}
G = \frac{r_{h} \left( 3r_{h} +\Lambda r_{h}^{3} + a \left( 3 -2 \Lambda r_{h}^{2} \right) \right)}{6 \left( 2r_{h} - a \right)},
\end{equation}
recovering  the result found   in \cite{17}.   The  globally stable thermodynamic system  can occur  if the Gibbs
free energy is negative $(G<0)$. However,  the unstable state  corresponds to   a positive Gibss 
free energy $(G>0)$.    To examine such behaviors, Fig.(\ref{7})    shows the variation of $ G$   as a function of $r_H$  for certain acceptable  regions of the reduced moduli space. 

Considering large radius values,  it  follows  that  this  function vanishes at   a particular $r_h= r^0_h$.  For   $r>r^0_h$,   the function $G$  takes negative values  revealing  that  the system is  stable. Otherwise, the  system  is globally unstable.

\subsection{Local stability}
\begin{figure}[h!]
    \centering
\begin{tabular}{cc}
\includegraphics[width=8cm,height=6.5cm]{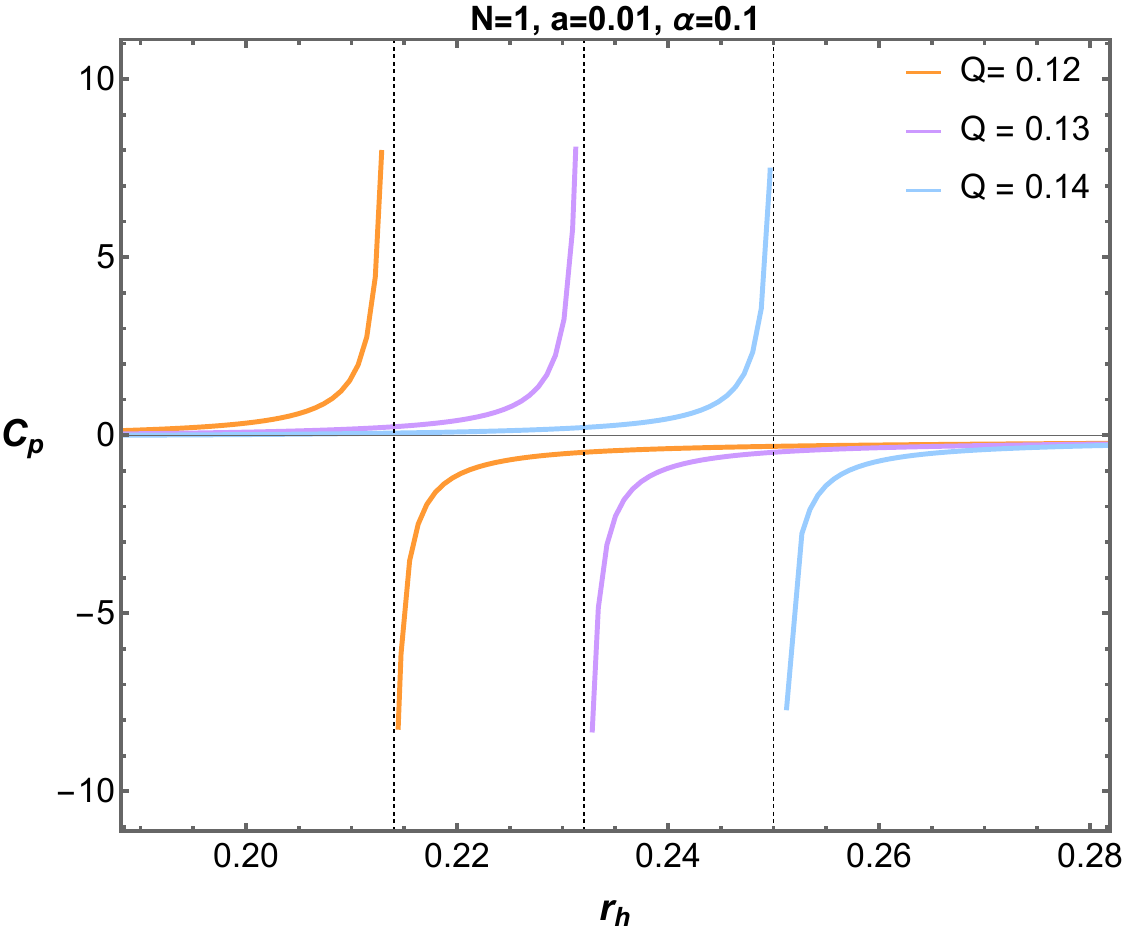} &
\includegraphics[width=8cm,height=6.5cm]{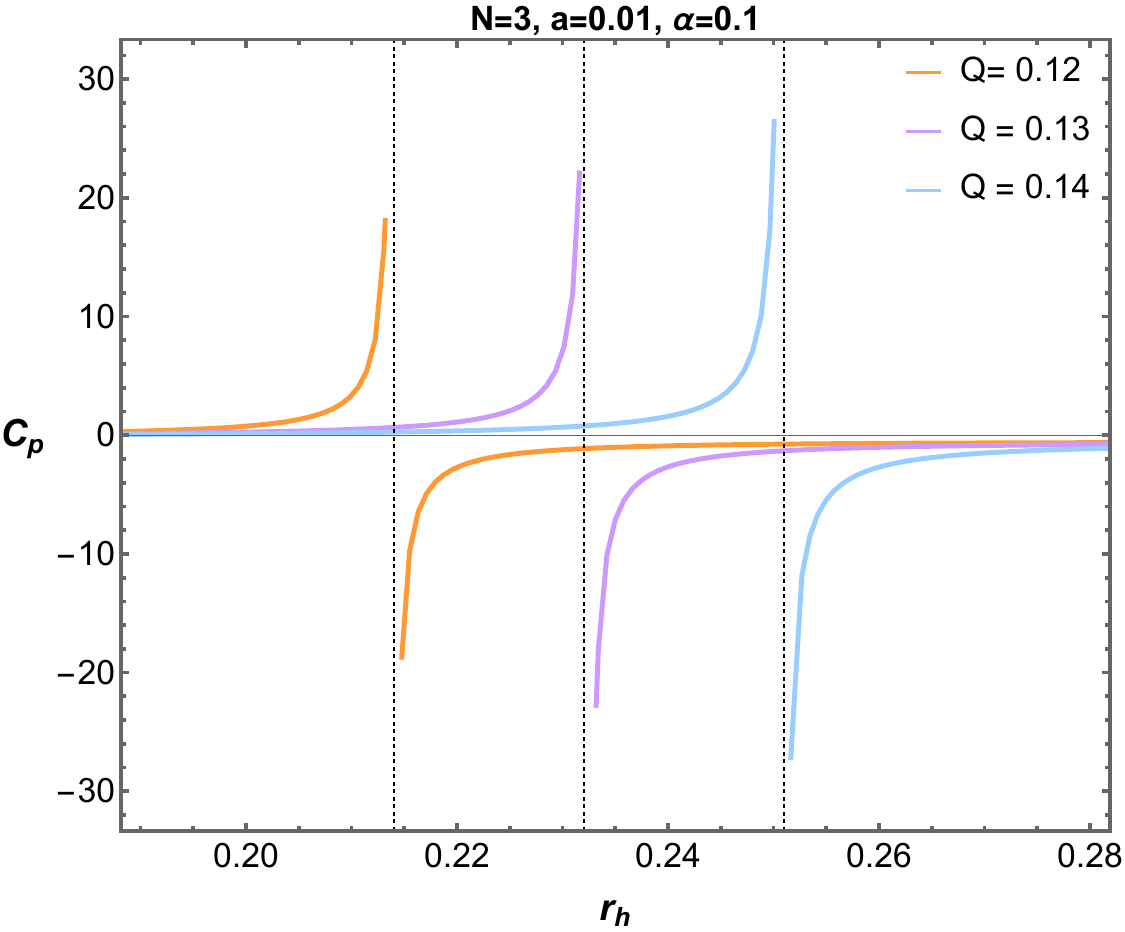}
\end{tabular}
\begin{tabular}{cc}
\includegraphics[width=8cm,height=6.5cm]{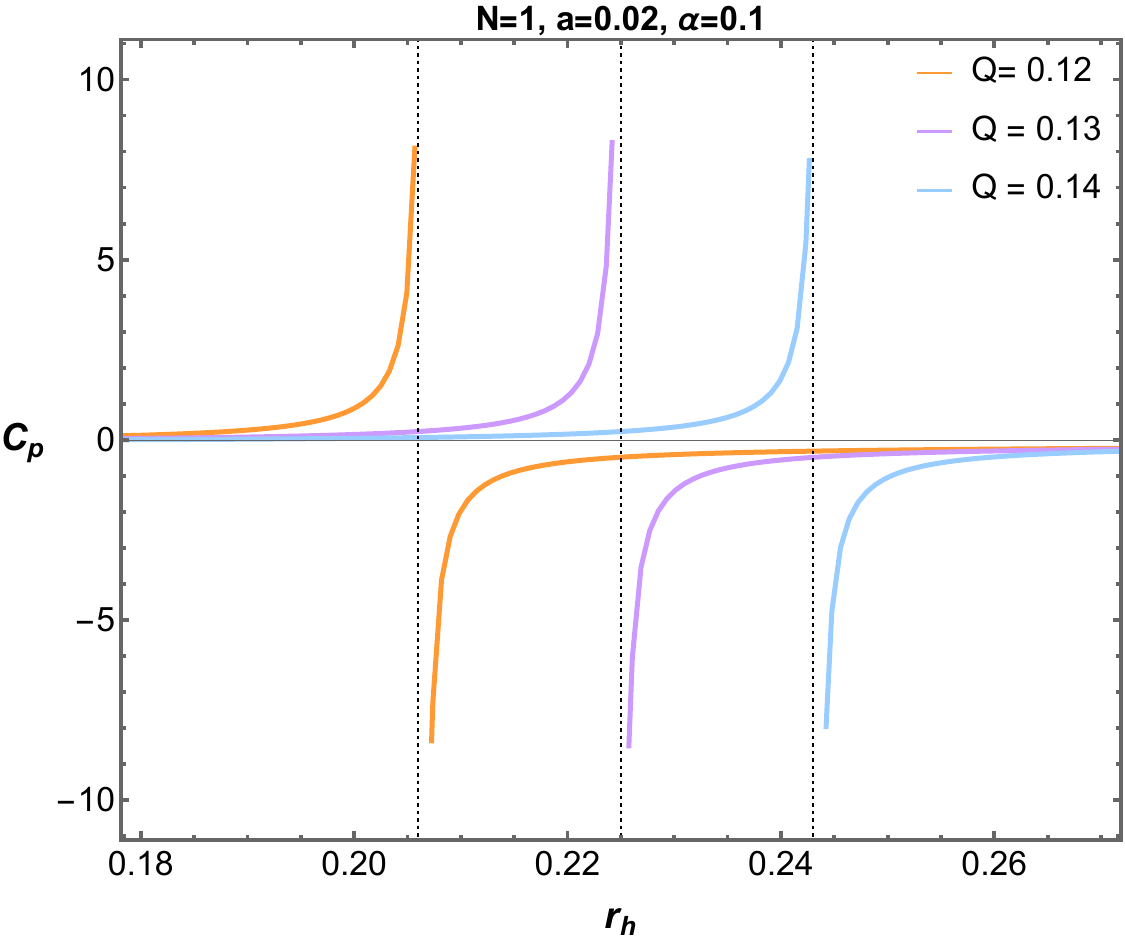} &
\includegraphics[width=8cm,height=6.5cm]{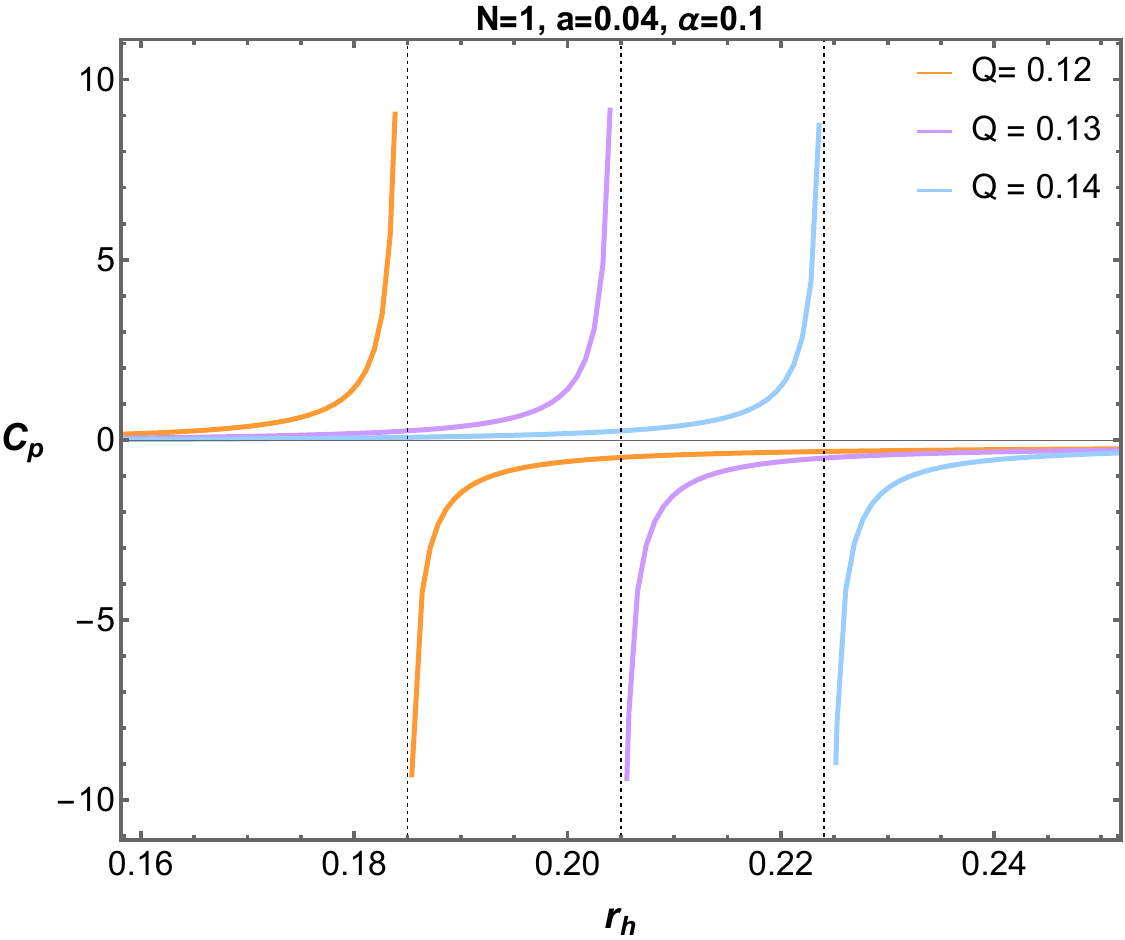}
\end{tabular}
\begin{tabular}{cc}
\includegraphics[width=8cm,height=6.5cm]{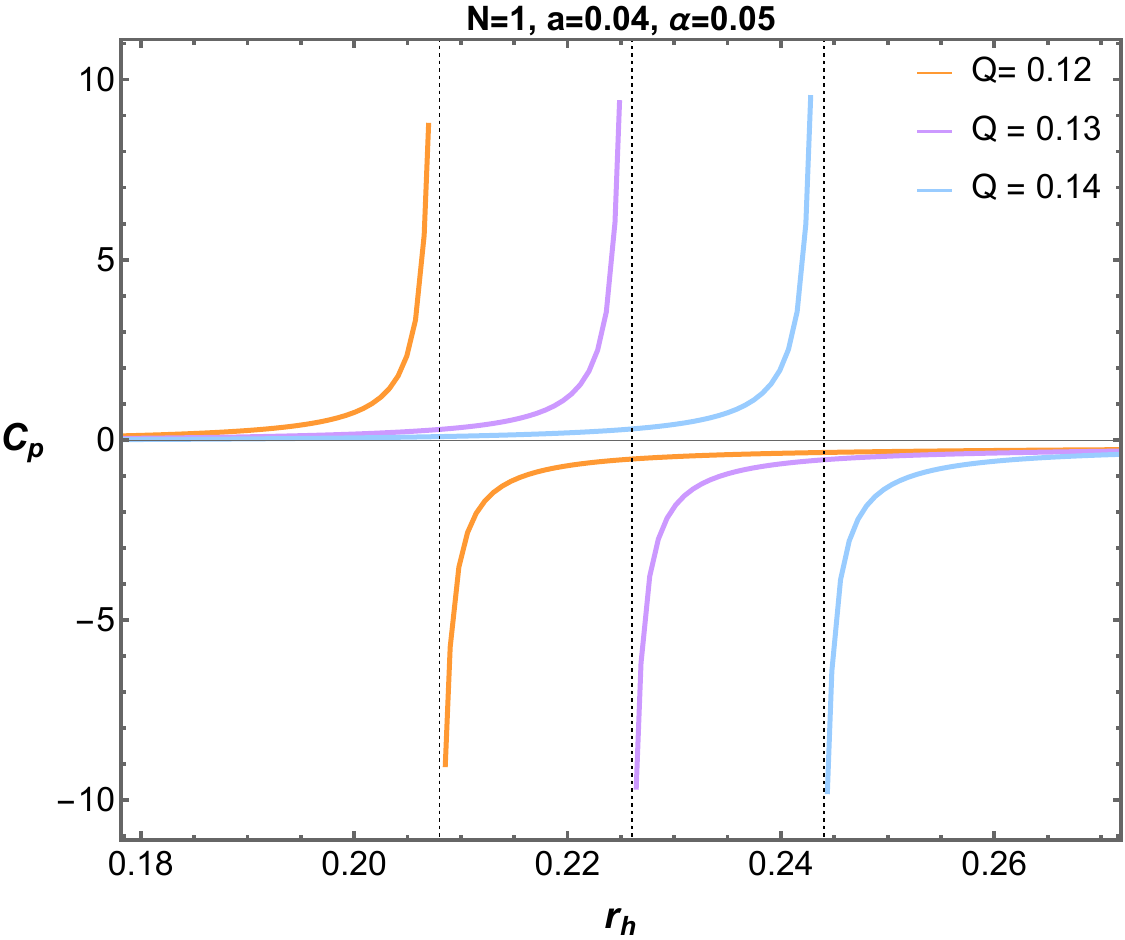} &
\includegraphics[width=8cm,height=6.5cm]{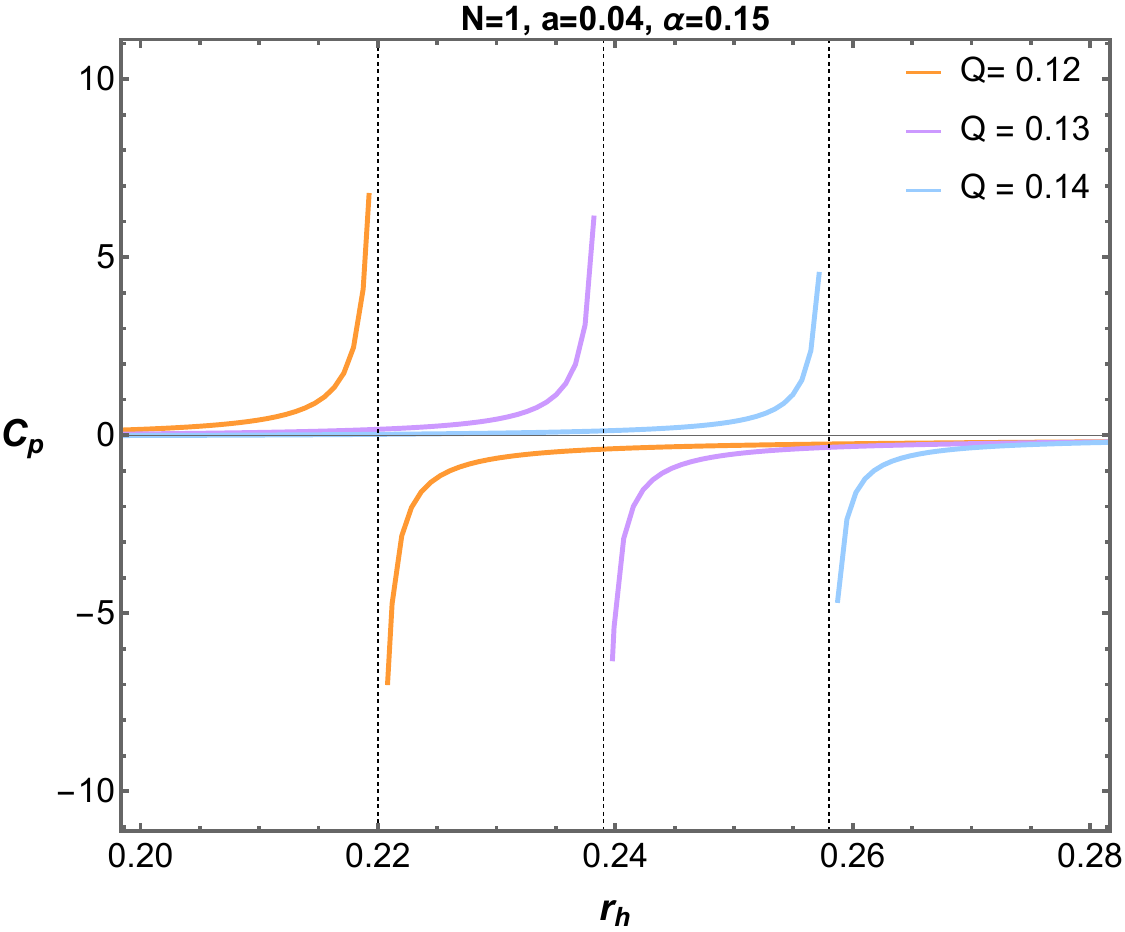}
\end{tabular}
    \caption{Heat capacity as a function of  $r_h$  for $\Lambda=-0.1$ and $w=-2/3$ by taking different values of the remaining parameters.}
    \label{4}
\end{figure}

In order to approach  the local stability, we need to compute the heat capacity using the relation    \begin{equation}
C_p = T_H  \frac{\partial S}{\partial T_H}.
\end{equation}
By help of the   Hawking temperature $T_H$ given in Eq.(\ref{xxx}), the heat capacity  is found to be 
{\footnotesize
\begin{equation}
C_p = \frac{4r_h^{3} \pi \left(
9N(2wr_h-(w-\dfrac{1}{3})a)+2\Lambda r_h^5(2a-3r_h)+6(\alpha-1)r_h^3(a-r_h)+3Q^2(2r_h^2-4ar_h+a^2)
\right) }{-9NDr_h^{2-3w} - 4 \Lambda r_h^{5}(a^{2}+3r_h^{2}-3ar_h) + 6r_h^{3}(\alpha-1)(2r_h^{2}-4ar_h+a^{2}) + 6Q^{2}(6r_h^{3}-18ar_h^{2}+11a^{2}r_h-6a^{3})}
\end{equation}}
where one has used 
\begin{equation}
D=12w(w-1)r_h^2+6ar_h(w^2+\dfrac{2}{3}w-\dfrac{5}{2})+(w-1)(4a^2+\alpha (a-2r_h)(3w-2)-8w(a-\dfrac{5r_h}{2}).
\end{equation}
Taking $Q=0$ and removing the external parameters, we  recover the following expression 
\begin{equation}
C_p =\frac{2 \pi  \,r_h^{2} \left(\Lambda  \,r_h^{2}-1\right)}{\Lambda  \,r_h^{2}+1},
\end{equation}
 representing the  commutative AdS--Schwarzschild black hole  reported in \cite{171}.  By using the sign of the heat capacity, we can determine the stability of the corresponding black hole solutions. In fact, a locally stable thermodynamic system can appear if $C_p>0$, while an unstable solution emerges if $C_p<0$. An illustration of this phenomenon is shown in Fig.(\ref{4}), in which $C_p$ is plotted as a function of $r_{h}$ for selected points in the reduced  moduli space.

For a specific point in the parameter space, we observe that the heat capacity curves become discontinuous at critical values $r_{h}=r_{h}^{c}$ corresponding to the minimum temperature. Fixing  the values of  $Q$,  $\alpha$, $N$ and $w$, we observe that $r_{h}^{c}$  decreases  by increasing   the NC parameter $a$.  Similar behaviors appear  for the  cloud string parameter  $\alpha$ and the   quintessence field parameter $N$. In this way, the external moduli space  affects  the  position   of  the heat capacity divergence   showing its effect on the thermodynamic stability. Furthermore, it has been pointed out that two distinct branches appear, indicating that the proposed models ensure a transition of the black hole from
a stable state to an unstable state, specified  by $r_h<r_{h}^{c}$ and $r_h>r_{h}^{c}$, respectively.

\section{Criticality  and  universality behaviors}

In this section, we   would like to approach certain    criticality  and  universality behaviors  of  the noncommutative RN--AdS black holes with a cloud of strings and quintessence dark energy fields.  At specific 
points of the   black hole moduli spaces, we show that  any critical  universal number  $X_N$ can
be split  as follows
\begin{equation}
X_N=X_0+NX,
\end{equation}
where $X_0$   is the  universal number of   the  noncommutative  RN--AdS black holes \cite{18}.  $X$  is an extra contribution  which depends on  the cloud of strings and the quintessence field parameters. To do so,  the first  step is to   establish    the equation of state.  The latter  is a fundamental aspect of thermodynamics, describing
the relationship between the   state variables. Similar to ordinary thermodynamic systems,  the black
hole systems may exhibit critical behaviors near phase transitions.  This  can play a crucial role in understanding and identifying critical phenomena. Moreover,
the variation of the entropy in terms of the temperature is relevant in the identification of
interesting universal quantities. To establish the equation of state, we use  the expressions associated with  the temperature and  the pressure.   In the extended phase space,  the cosmological constant $\Lambda$ is  considered as a thermodynamic variable being  the pressure
\begin{equation}
P = -\frac{\Lambda}{8\pi}.
\end{equation}
Such a thermodynamic description is not only more complete, but also encourages the emergence of rich phase structures and critical phenomena similar to those observed in the ordinary thermodynamic systems, such as Van der Waals fluids. This approach forms the basis of the equation of state used to verify the $P$-$V$ criticality of the system  under investigation \cite{18,171}.   After calculations, we find 

{\footnotesize
\begin{equation}
P =\frac{3 N r_h^{2-3 w} \! \left(\left(3 w -1\right) a -6 \mathit{wr_h} \right)+12 T r_h^4 \! \left(2 r_h -a \right)+6 r_h^3 \left(\alpha -1\right) \! \left(r_h -a \right)+3 Q^2 \! \left(a^{2}-4 a r +2 r^{2}\right)}{16 r^{5} \left(-2 a +3 r \right) \pi} \label{P}
\end{equation} }
Taking  $N=\alpha=w=0$, we recover the expression found in \cite{WM} being 
\begin{equation}
 P =\frac{12 \pi  T r_h^4 \! \left(2 r_h -a \right)+6 r_h^3 \! \left(a -r_h \right)+3 Q^2 \! \left(a^{2}-4 a r_h +2 r_h^{2}\right)}{16 r_h^{5} \left(3 r_h -2 a \right) \pi}.
\end{equation}
Vanishing the electric charge, we find the expression of  the Schwarzschild-AdS black hole in noncommutative geometry 
\begin{equation}
 P =\frac{3\left(2 \pi T a r_h - 4 \pi T r_h^{2} - a + r_h\right)}{8 r_h^{2} (2a - 3r_h)\pi},
\end{equation}
reported in   \cite{17}. Having established the equation of state, we move to investigate the universality  behaviors  by approaching the  $P$-$V$ criticality and the Joule-Thomson effect.
 \subsection{$P$-$V$ criticality behaviors } 
 To get certain universal aspect, we should determine   the thermodynamic critical  values. To establish  the associated expressions,   the black hole thermodynamic volume is needed.  This is found to be 
 \begin{equation}
 V=\frac{4 \pi  \,r_h^{3}}{3}.\label{V}
 \end{equation}
 Working out directly  the corresponding  quantities  can be considered is a highly non-trivial task as it requires  more reflective thinking.  However,   we can exploit  certain  techniques   explored in    \cite{WM,170,1700}.  Implementing  a new NC  parameter $s$ linked to $a$    as follows
  \begin{equation} 
\frac{2a}{3r_h}=s,
\end{equation}
the computations  become possible.  The critical values that we are after  could be  found  where certain conditions should be imposed on such a  new parameter in order to   obtain  acceptable quantities by  solving the conditions
\begin{equation}
\frac{\partial P}{\partial v }=0,\hspace{1.5cm}\frac{\partial ^{2}P}{
\partial v^{2}}=0.
\end{equation}
An examination shows  that the solution of these equations  can be  derived  by fixing the value of $w$.  Taking  $w=-2/3$, 
such values  are found to be 
{\footnotesize
\begin{eqnarray}
P_c &=& \frac{\left(3 s -2\right)^{2} \left(\alpha -1\right)^{2}}{48 \left(9 s^{2}-24 s +8\right) \left(1-s \right) \pi  \,Q^{2}} \\ 
T_{c} &=& \frac{\sqrt{6}\, \left(3 N \sqrt{6}\, Q \left(s -\frac{8}{9}\right) \sqrt{\left(3 s -2\right) \left(\alpha -1\right) \left(9 s^{2}-24 s +8\right)}+16 \left(s -\frac{2}{3}\right)^{2} \left(\alpha -1\right)^{2}\right)}{8 \sqrt{\left(3 s -2\right) \left(\alpha -1\right) \left(9 s^{2}-24 s +8\right)}\, \pi  \left(4-3 s \right) Q}
\\
v_{c} &=& \frac{\sqrt{6}\, \sqrt{\left(3 s -2\right) \left(\alpha -1\right) \left(9 s^{2}-24 s +8\right)}\, Q}{\left(3 s -2\right) \left(\alpha -1\right)}.
\end{eqnarray}}
The critical triple $(P_{c}, T_{c}, v_{c})$ provides  a ratio  $\chi_N= \frac{P_c v_c}{T_c}$. Using the small limit  approximations,  this ratio can be factorized as 
\begin{equation}
\chi_N= \chi_0+ N \chi,
\end{equation}
where one has
\begin{equation}
\chi_0=   \frac{3}{8}+ \frac{3 s}{32}, \qquad   \chi= \frac{9  \sqrt{6}\, Q}{16}+\frac{27  \sqrt{6}\, Q \alpha}{32}-\frac{9  \sqrt{6}\, Q s}{128}.
\end{equation}
 In the absence of a cloud of strings and  quintessence fields, we recover the universal behavior   $\chi_0$ of   the  noncommutative RN--AdS black holes  reported in \cite{WM}.
To approach  universal behaviors similar to the  Van der Waals fluid 
systems, restrict  conditions should be imposed on the  moduli space.   Taking, for instance,   the following charge value
\begin{equation}
Q=\frac{2 s \sqrt{6}}{9 N \left(s -12 \alpha -8\right)},
\end{equation}
we recover  the universal ratio
\begin{equation}
\chi_N =\frac{3}{8},
\end{equation}
where certain conditions on  the external  parameters should be required. This value is precisely identical to that of the Van der Waals fluid  standing  as a
universal number predicted for any  ordinary  charged RN-AdS black hole.   To support such a  discussion,  we illustrate  the subregions of  the reduced moduli space 
for which the  studied black hole behave as   Van der Waals fluids.
\begin{figure}[h!]
\centering
\includegraphics{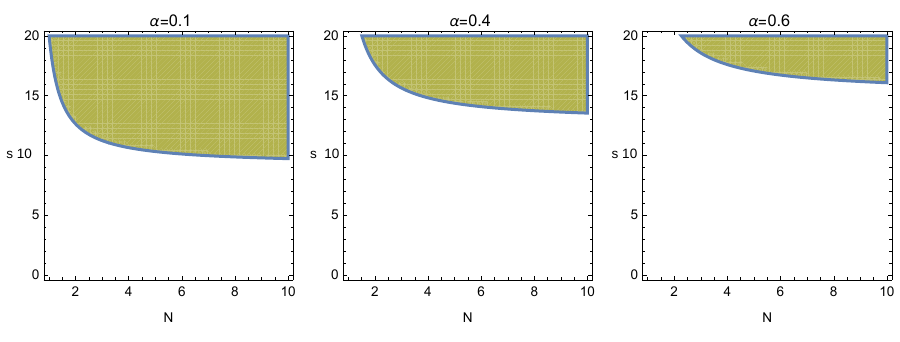} 
\caption{Allowed  subregions of  Van der Waals fluids  by varying $\alpha$.}
\label{055}
\end{figure}
It has been  remarked  that such regions are  relevant for small  values of the string cloud parameter $\alpha$.  More precisely, it has been  observed  that the size  of such regions decreases   with $\alpha$.  It follows that there are also certain subregions to which no  such  a behavior  is assigned. Their sizes increase 
with $\alpha$.

\begin{figure}[h!]
    \centering
\begin{tabular}{cc}
\includegraphics[width=8cm,height=6.5cm]{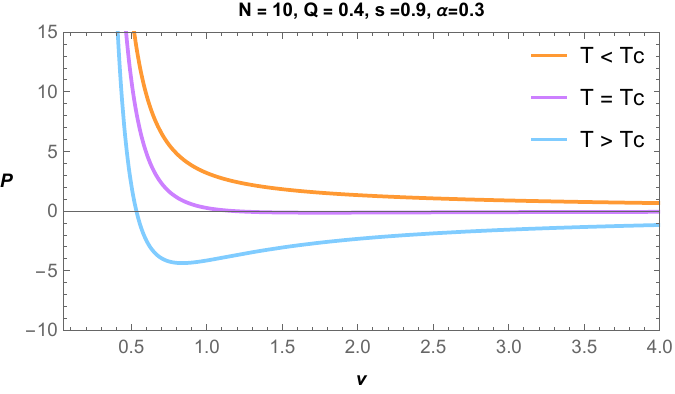} &
\includegraphics[width=8cm,height=6.5cm]{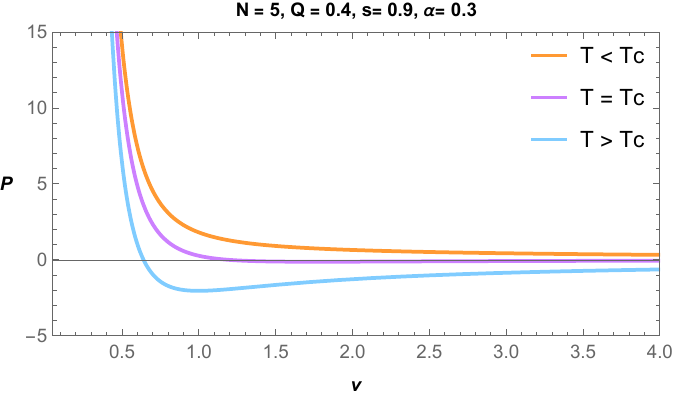}
\end{tabular}
\begin{tabular}{cc}
\includegraphics[width=8cm,height=6.5cm]{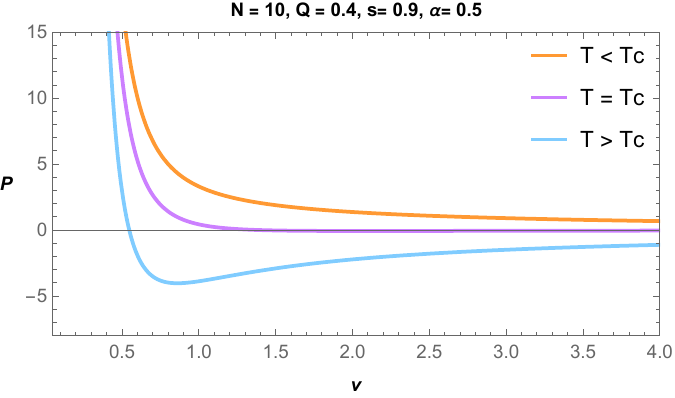} &
\includegraphics[width=8cm,height=6.5cm]{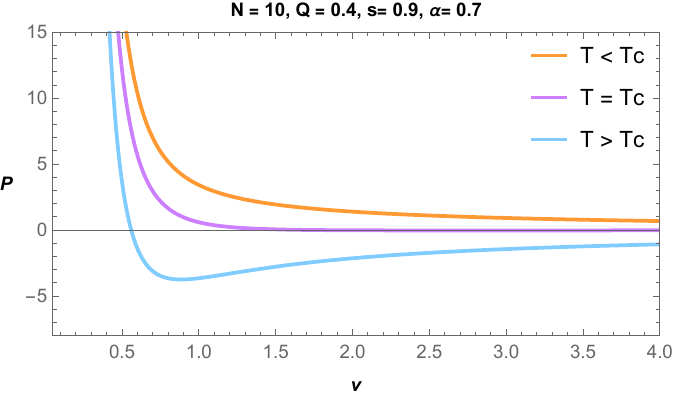}
\end{tabular}
\begin{tabular}{cc}
\includegraphics[width=8cm,height=6.5cm]{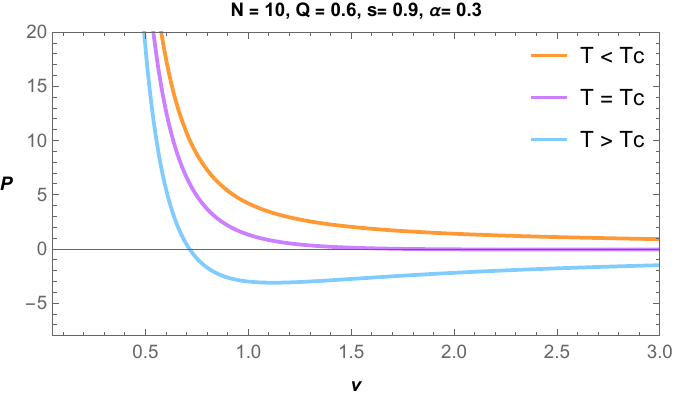} &
\includegraphics[width=8cm,height=6.5cm]{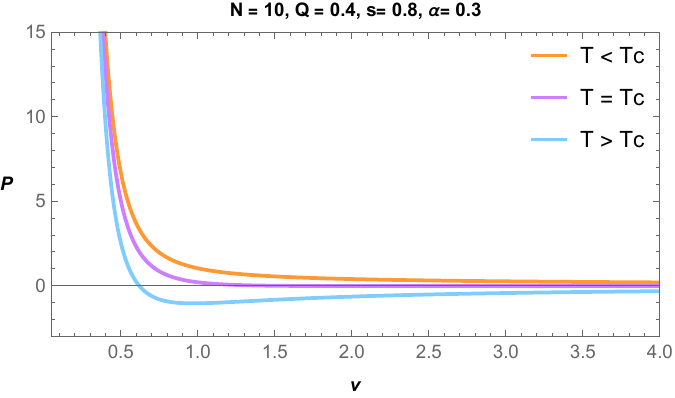}
\end{tabular}
\caption{Pressure in terms of $v$ with $w=-2/3$. }
\label{F41}
\end{figure}

 To reinforce this critical behavior analysis,    the $P$-$V$
diagram is illustrated in  Fig.(\ref{F41}) by varying the charge and the external parameters.

 For a temperature $T$ exceeding the critical value $T_c$, the system behaves like an ideal gas. The critical isotherm at $T = T_c$ is characterized by an inflection point at the critical pressure $P_c$ and  the critical volume $v_c$. For $T < T_c$, there is a thermodynamically unstable region. The $P$-$V$ diagram clearly resembles that of a Van der Waals fluid. Besides the charge, the  external parameters impact  the thermodynamic  criticality  of the system under examination. In particular, it has been  observed that they affect   the $P$-$V$ diagram structures providing relevant modifications.

\subsection{Joule-Thomson effect} 
The most common and classic physical process used to describe the change in temperature of a gas passing from a high-pressure section to a low-pressure section through a porous plug is referred to as Joule-Thomson expansion. This process essentially focuses on the mechanism of a gas expansion, which reflects the cooling effect and the heating effect  with the enthalpy remaining constant throughout the process. This process relays on  the Joule-Thomson coefficient, which   reads as 
\begin{equation}
\mu=\left( \dfrac{\partial T}{\partial P} \right)_{M}=\dfrac{1}{C_{P}} \left[ T \left( \dfrac{\partial V}{\partial T} \right)_{P}-V  \right]. \label{m}
\end{equation}
This can be exploited to  extract  extra  thermodynamical behaviors of  the black holes under examination\cite{21,22,23,24,25,27,28,29,30,31,32,320}. 
To obtain the temperature inversion,   such a  coefficient   should be computed.   To do so,   one has to  establish  the equation of state  as a function of the   thermodynamic volume. Using Eq.(\ref{V}), Eq.(\ref{P}),  and Eq.(\ref{xxx}),   taking  $w=-2/3$,  the   temperature   can be expressed  in terms of the  volume and the pressure.  It is found to be

\begin{equation}
T = \frac{24 P V  \left(\frac{6V}{\pi}\right)^{{1}/{3}} \left(s -1\right)-\left(3s -2\right)  \left(\frac{6V}{\pi}\right)^{{2}/{3}} \left(\alpha -1\right)+ Q^{2} \left(9s^{2}-24 s +8\right)}{6 V \left(3 s-4 \right)}
.\label{t}
\end{equation}
 Using Eq.(\ref{t}) and the second part of Eq.(\ref{m}), we can obtain  the temperature inversion by vanishing the   Joule-Thomson coefficient.  The   repeated inversion temperature $T_i$ is  shown  to be 
\begin{equation}
T_i = \frac{24 PV \,\left( \dfrac{6V}{\pi}\right) ^{{1}/{3}}    \left(s -1\right)  \left( \dfrac{6V}{\pi}\right) ^{{2}/{3}} \left(\alpha -1\right)\left(3s -2\right) -3 \,Q^{2}\left(9 s^{2}-24 s +8 \right) }{18 V\left(3   s -4\right)}.  \label{tt}
\end{equation}
Exploiting  the volume quantity,  this temperature   can be expressed as \begin{equation}
T_i=\frac{64 P \pi  \left(s -1\right) r^{4}+4 \left(\alpha -1\right) \left(3s -2\right) r^{2}-3 Q^{2} \left(9s^{2}-24 s +8\right)}{24 \pi  \,r^{3} \left(3 s-4 \right)}.
\label{ti1}
\end{equation}
 Eq.(\ref{t}) leads to 
\begin{equation}
T= \frac{64 P \pi  \left(s -1\right) r^{4}-4 \left(\alpha -1\right) \left(3s -2\right) r^{2}+ Q^{2} \left(9s^{2}-24 s +8\right)}{8 \pi  \,r^{3} \left(3 s-4 \right)}
\label{ti2}
\end{equation}
Subtracting Eq.(\ref{ti1}) form Eq.(\ref{ti2}),  we   find the following constraint
 \begin{equation}
64 P \pi  A \,r^{4}- 8B \,r^{2}+3 Q^{2} C = 0,   
 \end{equation}
where one has used 
\begin{equation}
\begin{aligned}
A=&s-1,\\
B=&\left(3s -2\right)\left(\alpha-1 \right), \\
C=& 9s^2-24s+8.
\end{aligned}
\end{equation}
 $P_{i}$   represents the inversion pressure.  The event horizon associated  with  the inversion temperature is found to be 
 \begin{equation}
r_i = \frac{\sqrt{3}\, \sqrt{P_i \pi  A \left(B -\sqrt{-12 P_i \pi  A \,Q^{2} C +B^{2}}\right)}}{4 P \pi  A}.
\end{equation}
By inserting this root into Eq.(\ref{ti1}),  we get  the expression of   the inversion temperature 
\begin{equation}
T_{i}=\frac{2 \left(8 P_i \pi  A \,Q^{2} C -B K\right) P A \sqrt{3}}{\sqrt{\pi}\, KL \sqrt{P A K}},
\end{equation}
where one used  $K=\left(B -\sqrt{-12 P_i \pi  A \,Q^{2} C +B^{2}}\right)$ and $L= 4-3s$.  Vanishing   the value of $P_i$,  the inversion temperature reaches its minimum value
 \begin{equation}
T_{i}^{min}=\frac{\sqrt{6}\, B^{2}}{9\pi  QL\sqrt{BC} }. 
 \end{equation}
It  is worth noting   that, in  charged black hole physics,  the minimum inversion and  the critical temperatures generate  a  ratio  expressed as  \begin{equation}\xi_N= \dfrac{T_{i}^{min}}{T_{c}}. \end{equation}
 Taking  small external parameters,  the  computations leads  to 
\begin{equation}
\xi_N=\xi_0+N\xi,
 \end{equation}
where  we have 
 \begin{equation}
\xi_0=\frac{1}{2}, \qquad  \xi= \frac{3  \sqrt{6}\, Q}{4}+\frac{9  \sqrt{6}\, Q \alpha}{8}-\frac{9  \sqrt{6}\, Q s}{32}.
 \end{equation}
It is denoted that  $\xi_0=\frac{1}{2}$   holds 
for   the  noncommutative  RN--AdS black holes \cite{WM}.  In the presence of a cloud of strings and  quintessence fields,    the  universal number $\xi_N=\frac{1}{2}$   can be recovered   by imposing the following  constraint on the internal moduli space coordinates 
 \begin{equation}
s=\frac{8}{3}+4 \alpha
\end{equation}
for any charge value.  This constraint not only recovers the  universal ratio but also it   reduces the number of the external  parameters.       The  isenthalpic curves and inversion curves can be useful in supplementing the discussion. In this way, we can identify the region where the constant enthalpy curve has a higher slope than that of the inverse curve.

\begin{figure}[h!]
\centering
\begin{tabular}{cc}
\includegraphics[width=8cm,height=6.5cm]{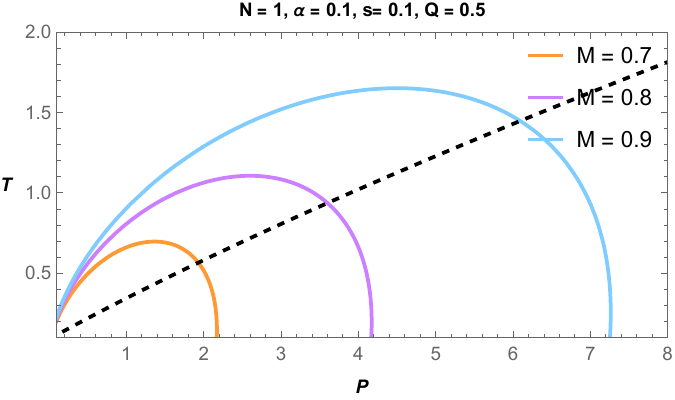} &
\includegraphics[width=8cm,height=6.5cm]{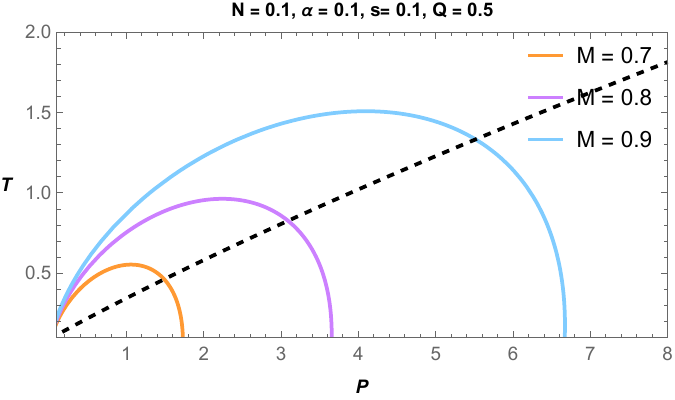}
\end{tabular}
\begin{tabular}{cc}
\includegraphics[width=8cm,height=6.5cm]{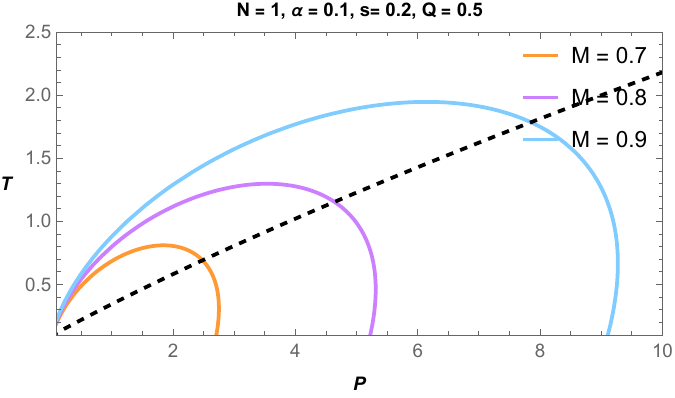} &
\includegraphics[width=8cm,height=6.5cm]{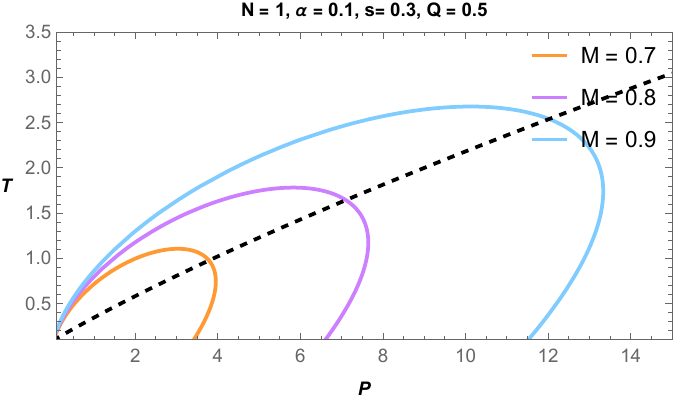}
\end{tabular}
\begin{tabular}{cc}
\includegraphics[width=8cm,height=6.5cm]{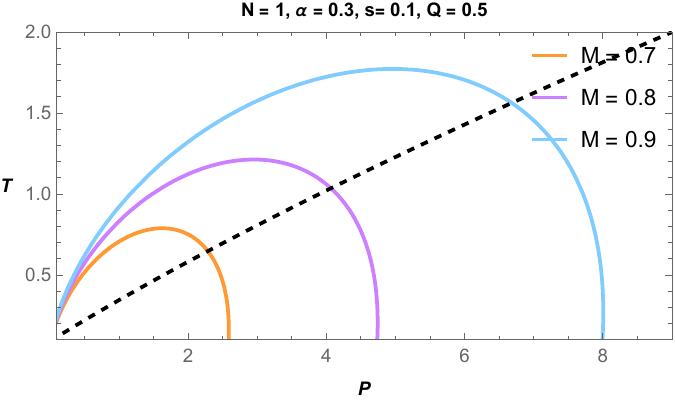} &
\includegraphics[width=8cm,height=6.5cm]{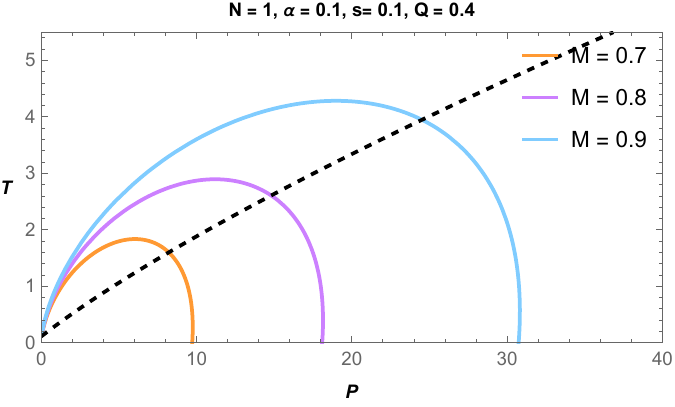}
\end{tabular}
\caption{Inversion (dashed lines) and isenthalpic (solid lines) curves for noncommutative RN-AdS black holes.}
\label{55}
\end{figure}

This could provide  the  region  where  the cooling  occurs.  The  sign  of  the  slope  of  the  isenthalpiccurves changes under the inversion curves  showing  that this  region  presents  signs  of  warming.  Indeed, Fig.(\ref{55}) illustrates  the inversion curves dividing  the $(T, P)$ diagram into two distinct zones. Above the inversion curves, the system cools, while below them, it warms. 
This can be seen from the slope of the isenthalpic curves. As for the inversion curve itself, there is neither warming nor cooling. Moreover,  the boundary between the two regimes is marked by an inversion curve.

\newpage

\section{Deflection angle computations}
 In this section,  we investigate certain optical  properties  of  by noncommutative quintessential RN–AdS black holes in the presence of a cloud of strings.  Since we consider the non-rotating case,  we limit the  analysis to  the deflection of light rays. It is denoted that the  deflection angle is evaluated using the Gauss–Bonnet theorem. This provides a global and geometrically elegant method for computing  the  weak gravitational lensing within the optical geometry framework, as originally proposed by Gibbons and Werner \cite{A117,A118,A119}.  In the scenario under consideration, the combined effects of the non-commutativity, the quintessence, and the string cloud provide a richer description, which is likely to generate observable deviations from the standard predictions of the RN–AdS model.  Considering that both the observer ($R$) and the source ($S$) are located at finite distances on the equatorial plane, The expression for the angle of deviation can be stated as follows
 \begin{equation}
\Theta=\Psi_{R}-\Psi_{S}+\phi_{SR}
\label{a3}
\end{equation}
where  $\Psi_{R}$ and $\Psi_{S}$ represent the angles between the light rays and the radial direction at the positions of the observer and the source, respectively.  The angle $\phi_{SR}$ indicates  the longitudinal separation between the source and the observer, as introduced in \cite{A117}. Accordingly, the separation angle can be  expressed as 
\begin{equation}
\phi_{RS} = \int^R_S d\phi= \int^{u_0}_{u_S}\frac{1}{\sqrt{F(u)}}du +\int^{u_0}_{u_R}\frac{1}{\sqrt{F(u)}}du
\end{equation} 
where $u_{S}$ and $u_{R}$ are the inverse of the distances from the black hole to the source and the observer, respectively. The parameter $u_{0}$   denotes the inverse of the closest approach distance $r_{0}$.  $b$  denotes   the  impact parameter $\frac{L}{E}$.  In this way,  the  function $F(u)$ is formulated as follows
\begin{eqnarray}
F(u)=\left(\frac{1}{u^2} \frac{du}{d\phi}\right)^2. 
\end{eqnarray}
Dealing with  the metric of the  noncommutative quintessential RN–AdS black holes in the presence of a cloud of strings for $\omega=-\frac{2}{3}$ and taking the order $\mathcal{O}(M,\Lambda,a,\alpha,N,Q^2)$, 
the computations provide 
\begin{equation}
F(u)=-a M u^4+a Q^2 u^5+\frac{1}{b^2}+\frac{\Lambda }{3}+2 M u^3+N u-Q^2 u^4+\alpha  u^2-u^2+\mathcal{O}(M,\Lambda,a,\alpha,N,Q^2).
\end{equation}
According to  the algorithm developed in \cite{A117,A118,A119} to obtain the $\Psi$ terms, we  should first  derive the  expression for $\sin(\Psi)$  by taking a fixed value of $w$. Taking $\omega=-\frac{2}{3}$,  we get 
\begin{equation}
\sin(\Psi)=\frac{b u^2 }{\sqrt{3}u^2}\sqrt{3 u^4 \left(a M+Q^2\right)-3 a Q^2 u^5-\Lambda -6 M u^3-3 N u-3 (\alpha -1) u^2}
\end{equation}
To establish  the expression of the deflection angle, a further expansion is required. This may  lead  to lengthy expressions for $\phi_{SR}$ and $\Psi_{S}-\Psi_{R}$. Following the well-known steps and considering the limits $u_S \ll 1$ and $u_R \ll 1$, we finally arrive at the  expression form  for the deflection angle
\begin{equation}
\Theta=\Theta_{\text{RN-AdS}}+\Theta_{(\alpha,a)}+N \Theta_N 
\end{equation}
By taking the  orders $\mathcal{O}(M,\Lambda,a,\alpha,N,Q^2)$,  we obtain  
{\footnotesize
\begin{eqnarray}
\Theta_{\text{RN-AdS}}=-\frac{32 M Q^2}{3 b^3}-\frac{3 \pi  Q^2}{4 b^2}-\frac{b \Lambda  M}{3}+\frac{8 \Lambda  M Q^2}{b}+\frac{4 M}{b}-\frac{1}{6} b \Lambda  \left(\frac{1}{{u_R}}+\frac{1}{{u_S}}\right)-2 \pi  \Lambda  Q^2.
\end{eqnarray}}
The second term  is found to be 
{\footnotesize
\begin{eqnarray}
\Theta_{(\alpha,a)}&=& \frac{2835 \pi  a \alpha  M Q^2}{64 b^4}+\frac{315 \pi  a M Q^2}{32 b^4}-\frac{8 a \alpha  Q^2}{b^3}-\frac{8 a Q^2}{3 b^3}-\frac{15 \pi  a \alpha  M}{8 b^2}+\frac{945 \pi  a \alpha  \Lambda  M Q^2}{32 b^2}+\frac{105 \pi  a \Lambda  M Q^2}{16 b^2}\nonumber\\&-&\frac{3 \pi  a M}{4 b^2}-\frac{4 a \alpha  \Lambda  Q^2}{b}-\frac{4 a \Lambda  Q^2}{3 b}-\frac{5}{8} \pi  a \alpha  \Lambda  M-\frac{1}{4} \pi  a \Lambda  M+\frac{\pi  \alpha }{2}+\frac{64 \alpha  M Q^2}{b^3}\\&-&\frac{\pi  \alpha  Q^2}{8 b^2}-\frac{11}{6} \alpha  b \Lambda  M-\frac{8 \alpha  M}{b}+\frac{32 \alpha  \Lambda  M Q^2}{b}-\frac{\alpha  b \Lambda }{12 }\left(\frac{1}{{u_R}}+\frac{1}{{u_S}}\right)-\frac{5}{8} \pi  \alpha  \Lambda  Q^2. \nonumber
\end{eqnarray}}
The last one is shown to be 
{\footnotesize
\begin{eqnarray}
\Theta_N &=&\alpha  Q^2 \left(\frac{105 \pi  a}{16 b^2}+\frac{12}{b}\right)+Q^2 \left(\frac{15 \pi  a}{8 b^2}+\frac{4}{b}\right)+\alpha  M Q^2 \left(-\frac{480 a}{b^3}-\frac{945 \pi }{16 b^2}\right)+M Q^2 \left(-\frac{96 a}{b^3}-\frac{105 \pi }{8 b^2}\right)+\frac{21}{8} a \alpha  b \Lambda  M\nonumber \\&+&\alpha  M \left(\frac{12 a}{b}+\frac{15 \pi }{4}\right)+\frac{11}{12} a b \Lambda  M+\alpha  \Lambda  M Q^2 \left(-\frac{240 a}{b}-\frac{315 \pi }{16}\right)-\frac{48 a \Lambda  M Q^2}{b}+M \left(\frac{4 a}{b}+\frac{3 \pi }{2}\right)\nonumber\\&+&\alpha  \Lambda  Q^2 \left(\frac{35 \pi  a}{16}+\frac{21 b}{8}\right)+\Lambda  Q^2 \left(\frac{5 \pi  a}{8}+\frac{11 b}{12}\right)-\frac{3 \alpha  b}{2}-\frac{1}{8} \left(\frac{1}{{u_R}}+\frac{1}{{u_S}}\right) (5 \alpha  b \Lambda  M) \\&-&\frac{1}{4} \left(\frac{1}{{u_R}}+\frac{1}{{u_S}}\right) (b \Lambda  M)-\frac{1}{8} \left(\frac{1}{{u_R}^2}+\frac{1}{{u_S}^2}\right) \alpha  b \Lambda -\frac{ b \Lambda }{12} \left(\frac{1}{{u_R}^2}+\frac{1}{{u_S}^2}\right)-2 b. \nonumber
\end{eqnarray}}

\begin{figure}[!ht]
		\begin{center}
		\centering
			\begin{tabbing}
			\centering
			\hspace{6.cm}\=\kill
			\includegraphics[scale=0.28]{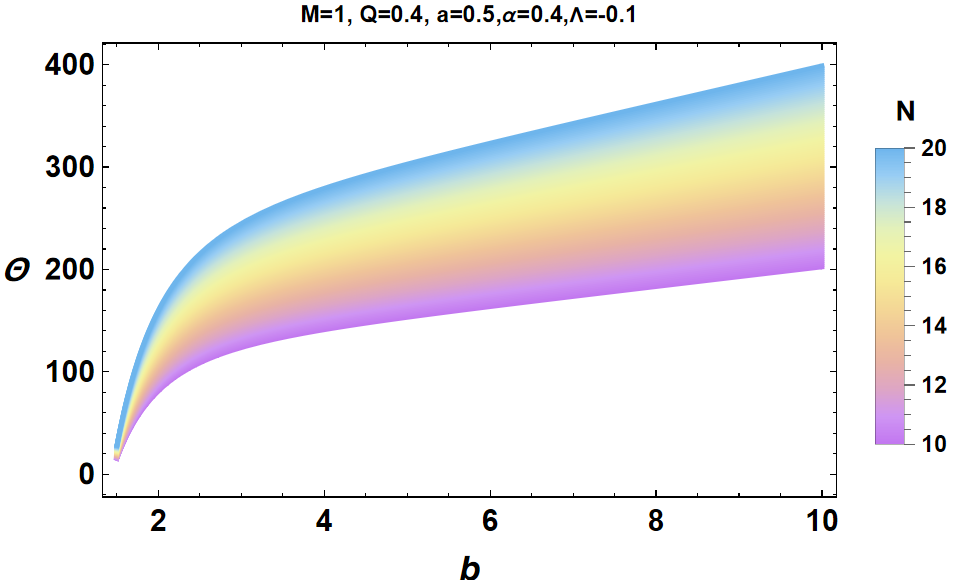} 
	\hspace{0.1cm}		\includegraphics[scale=0.28]{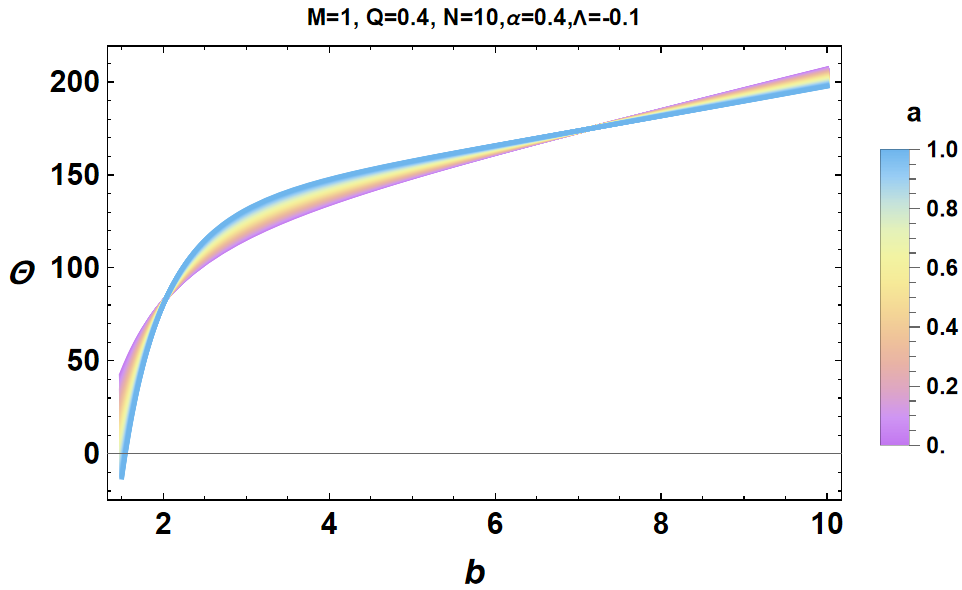}\hspace{0.1cm}	\includegraphics[scale=0.28]{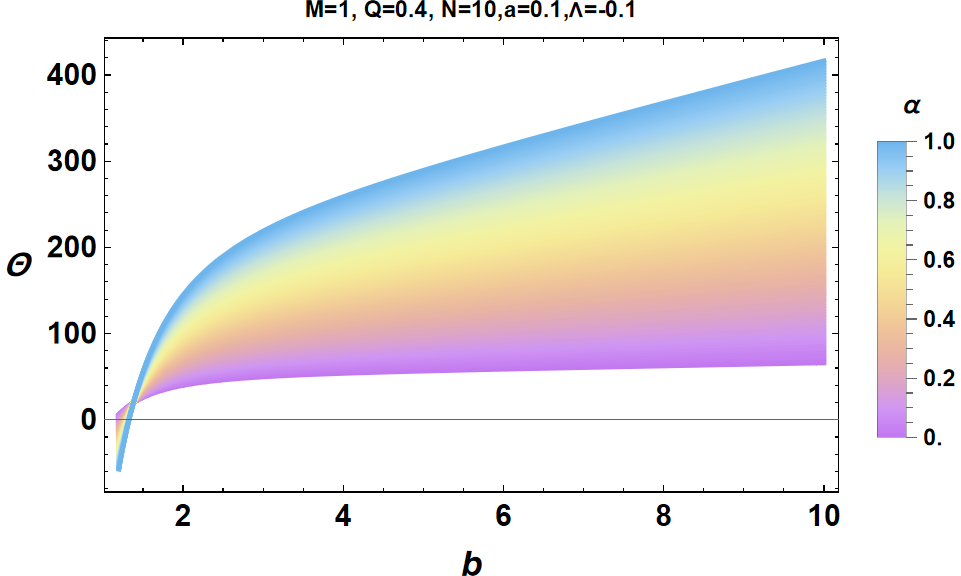}\\ 
	
		   \end{tabbing}
\caption{  \footnotesize  Deflection angle behaviors in terms of the impact parameter by varying  the external  parameters.}
\label{rsa}
\end{center}
\end{figure}

To examine the obtained expression, the  variation of  the deflection angle 
 is illustrated in Fig.(\ref{rsa}).
  Fixing the internal parameters,  the deflection angle  increases
by  augmenting  $b$.  Moreover, such an optical quantity increases with the parameter $N$, while for the external parameters $\alpha$ and $a$ it exhibits an oscillatory behavior.

\section{Conclusions}
In the presence of  a cloud of strings and 
dark energy  fields, we have studied the  Reissner--Nordström--AdS black holes in a NC spacetime with Lorentzian-smeared distributions. 
 In the context of such a deformed spacetime, we have investigated the thermodynamical  and the optical properties of charged AdS black holes. Concerning the thermodynamics,   we have focused first  on both local and global stabilities  and  critical behaviors. For  optical properties, we have investigated  the deflection angle of lights near such black holes. Concretely,  we have started by  analyzing the  global and the stability behaviors. In particular, we  have computed   the  Gibbs free energy  and  the  heat capacity  in order to identify  the regions where the black holes remain stable. Moreover, by relating the noncommutative parameter $a$ to the horizon radius $r_h$ through a constant parameter $s$, we have examined the $P$--$V$ criticality. Precisely, we have   derived  the critical pressure $P_c$, the critical  temperature $T_c$, and the critical specific volume $v_c$ in terms of the charge $Q$  and the external  parameters  $(a, \alpha,  N,  w)$. We have approached  an  universal  ratio $\chi_N=\frac{P_c v_c}{T_c}$. In the small-limit regime of the external parameters, we have recovered behaviors analogous to those of Van der Waals fluids by imposing certain constraints on the involved parameters. Additionally, we have investigated the Joule--Thomson expansion by computing  the universal ratio $\xi_N= \dfrac{T_{i}^{min}}{T_{c}}$. It has been shown that this  highlights both similarities and differences with Van der Waals systems, which further supports the validity of the proposed noncommutative black hole metric.

 Finally, we have   examined the dependence of such a quantity in terms of external  black hole
parameters. Concretely, we have  observed that the behavior of
the deflection angle is a  increasing function of  external parameters.

 This work  leaves certain open questions. Rotating solutions  could be a
possible extension of the present paper in order to approach shadow behaviors and make contact with empirical investigations including the findings of  EHT collaboration. We hope address such a  question in future works.

  {\bf Data availability}\\
  Data sharing is not applicable to this article.

\section*{Acknowledgements}
 M. J   gratefully acknowledges the financial support  of the CNRST in the frame of the PhD Associate Scholarship Program PASS.


\begin{thebibliography}{99}


\bibitem{1}
N. Seiberg, E. Witten, \textit{String Theory and Noncommutative Geometry}, JHEP 9909 032
 (1999), \texttt{ arXiv:hep-th/9908142}.
\bibitem{001} L.  Pauling,  E. Bright Wilson,    \textit{Introduction to Quantum Mechanics},    McGraw-Hill Inc.,US, 1935.

\bibitem{up12}
D. Kastor, S. Ray, J. Traschen, \textit{Smarr Formula and an Extended First Law for Lovelock Gravity}, Class.Quant.Grav. 27   235014 (2010), \texttt{arXiv:1005.5053}. 

\bibitem{up15}
D. Kastor, S. Ray, J. Traschen, \textit{Extended First Law for Entanglement Entropy in Lovelock Gravity,}  Entropy 18  6 212 (2016), \texttt{	arXiv:1604.04468}.
\bibitem{up11}
D. Kastor, S. Ray, J. Traschen, \textit{Enthalpy and the Mechanics of AdS Black Holes,} Class. Quant. Grav. 26 195011 (2009), \texttt{	arXiv:0904.2765}.

\bibitem{up13}
D. Kastor, S. Ray, J. Traschen, \textit{Mass and free energy of Lovelock black holes,} Class. Quantum Grav. 28 195022 (2011), \texttt{arXiv:1106.2764 [hep-th]}.

\bibitem{up14}
D. Kastor, S. Ray, J. Traschen, \textit{Chemical Potential in the First Law for Holographic Entanglement Entropy,}  J. High Energ. Phys.  120 (2014), \texttt{arXiv:1409.3521}.

\bibitem{up16}
D. Kastor, S. Ray, J. Traschen, \textit{Black Hole Enthalpy and Scalar Fields,} Class. Quantum Grav. 36 024002 (2019), \texttt{	arXiv:1807.09801}.
\bibitem{2}
 M. R. Douglas, N. A. Nekrasov,  \textit{Noncommutative field theory},  Reviews of Modern Physics, 73(4) 977 (2001).

 \bibitem{up9}
 E. Battista, \textit{Quantum Schwarzschild geometry
in effective-field-theory models of gravity}, Phys.Rev.D 109, 2, 026004 (2024), \texttt{	arXiv:2312.00450}. 

\bibitem{212}
A. Karch, B. Robinson, \textit{Holographic Black Hole Chemistry}, 	JHEP12(2015)073, \texttt{arXiv:1510.02472}.

\bibitem{2121}
R. Mancilla, \textit{Generalized Euler Equation from Effective Action:
Implications for the Smarr Formula in AdS Black
Holes}, \texttt{arXiv:2410.06605}.

\bibitem{adilsaidi}
A. Belhaj, M. Hssaini, E. L. Sahraoui, E. H. Saidi, \textit{Explicit Derivation of Yang-Mills Self-Dual Solutions on non-Commutative Harmonic Space},  Class.Quant.Grav. 18, 2339-2358  (2001),   \texttt{arXiv:hep-th/0007137}.
\bibitem{3}
D. Berenstein, R. G.  Leigh, \textit{Non-commutative Calabi–Yau manifolds,} Phys. Review Letters, 84(20) 4737–4740 (2000).

\bibitem{adil1}  A. Belhaj, E. H. Saidi,  \textit{On Non Commutative Calabi-Yau Hypersurfaces}, 
  Phys.Lett. B523, 191-198  (2001),    \texttt{arXiv:hep-th/0108143}. 

\bibitem{adil2}   A. Belhaj, J. J. Manjarin, P. Resco, \textit{On Non-Commutative Orbifolds of K3 Surfaces},  J.Math.Phys. 44, 2507-2520  (2003),    \texttt{arXiv:hep-th/0207160}.

\bibitem{adil3}   A. Belhaj, J.  Rasmussen, E.  H. Saidi, A.  Sebbar, \textit{Non-commutative ADE geometries as holomorphic wave equations},   Nucl.Phys. B727,  499-512  (2005),   \texttt{arXiv:hep-th/0504049}.


\bibitem{4}
P. Nicolini, A. Smailagic, E. Spallucci,  \textit{Noncommutative geometry inspired Schwarzschild black hole}. Physics Letters B, 632(4) 547–551 (2006).

\bibitem{45}
M. R. Douglas, C. M. Hull, \textit{  D-branes and the noncommutative torus, } Journal of High Energy Physics, 1998(02) 008 (1998).

\bibitem{5}
P. Nicolini, \textit{Noncommutative Black Holes, The Final Appeal To Quantum Gravity: A Review, } International Journal of Modern Physics A, 24(07), 1229–1308 (2009).

\bibitem{6}
F. Rahaman, A. Banerjee, I. Radinschi, M. Kalam, S. Islam, \textit{ Noncommutative geometry inspired charged black holes,} Physics Letters B, 694(1) 10–15 (2010).

\bibitem{66}
K. Nozari, S. H. Mehdipour, \textit{Hawking Radiation as Quantum Tunneling from Noncommutative Schwarzschild Black Hole}, Class. Quant. Grav. 25, 175015 (2008), \texttt{	arXiv:0801.4074}.

\bibitem{77}
K. Nozari, S. Islamzadeh, \textit{Tunneling of massive and charged particles from noncommutative Reissner-Nordström black hole},  Astrophys. Space Sci. 347 (2013) 299, \texttt{	arXiv:1207.1177}.


\bibitem{100}
K. Nozari, B. Fazlpour, \textit{Reissner-Nordström Black Hole Thermodynamics in Noncommutative Spaces},   	ActaPhys.Polon.B39:1363 (2008), \texttt{		arXiv:gr-qc/0608077}.

\bibitem{1001}
K. Nozari, S H. Mehdipour, \textit{Noncommutative inspired Reissner–Nordström black holes in large extra dimensions},  	Commun. Theor. Phys. 53, 503-513 (2010),  \texttt{	arXiv:0707.1080}.
 

\bibitem{up17}
B. L. Liu, Y. Zhang, Q. Q. Li,  \textit{Thermodynamics and P-v criticality of RN-AdS black hole surrounded by PFDM on the EGUP framework,} \texttt{arXiv:2509.08005}.
 
 
\bibitem{7}
W. Kim, E. J. Son, M. Yoon, \textit{Thermodynamic similarity between the noncommutative Schwarzschild black hole and the Reissner–Nordström black hole}, JHEP 04  042(2008), \texttt{arXiv:0802.1757}.

\bibitem{8}
P. Nicolini, \textit{ How strings can explain regular black holes},   Regular Black Holes: Towards a New Paradigm of Gravitational Collapse, Springer Series in Astrophysics and Cosmology. Springer, Singapore,  \texttt{arXiv:2306.01480}.

\bibitem{A7}
M. Bežanić, M. D. Ćirić, N. Konjik, B. Nikolić, A. Samsarov, \textit{Noncommutative fields in Reissner-Nordström black hole background,} (2025), \texttt{arXiv:2505.06181}.

\bibitem{9}
S. W. Wei, P. Cheng, Y. Zhong, X. N. Zhou,  \textit{Shadow of noncommutative geometry inspired black hole,}  Cosmology and Astroparticle Physics (JCAP), 08 004 (2015), \texttt{arXiv:1501.06298}.

\bibitem{10}
J. A. V. Campos, M. A. Anacleto, F. A. Brito, E. Passos, \textit{Quasinormal modes and shadow of noncommutative black hole},  Scientific Reports (2022), \texttt{arXiv:2103.10659}.
\bibitem{up10}
Z. L. Wang, E. Battista, \textit{Dynamical features and shadows of quantum Schwarzschild black hole in effective field theories of gravity,}  Eur.Phys.J.C 85 3, 304 (2025),  \texttt{	arXiv:2501.14516}.



\bibitem{ref3} 
S. G. Ghosh, \textit{Noncommutative geometry inspired Einstein-Gauss-Bonnet black holes,}  Class.Quant.Grav. 35  8 085008 (2018), \texttt{arXiv:1707.08174}.

\bibitem{WM}  W.  El Hadri, M.  Jemri, \textit{Thermodynamics and Criticality of Noncommutative RN-AdS Black Holes},  \texttt{arXiv:2509.00926}.



\bibitem{11}
 K. Akiyama and al., \textit{First M87 Event Horizon Telescope
Results. IV. Imaging the Central Supermassive Black Hole}, Astrophys. J.
\textbf{L4} (1) 875 (2019), \texttt{arXiv:1906.11241}.

\bibitem{12}
 K. Akiyama and al., \textit{First M87 Event Horizon Telescope
Results. V. Imaging the Central Supermassive Black Hole}, Astrophys. J.
\textbf{L5} (1) 875 (2019).

\bibitem{13}
 K. Akiyama and al., \textit{First M87 Event Horizon Telescope
Results. VI. Imaging the Central Supermassive Black Hole}, Astrophys. J.
\textbf{L6} (1)875 (2019).

\bibitem{14}
 F. Ahmed, A. R. P. Moreira, A. Bouzenada, \textit{Noncommutative Geometry Inspired AdS Black Hole with a Cloud of Strings
 Surrounded by Quintessence-like fluid}, \texttt{arXiv:2508.00740}.


\bibitem{15}
G. Mascher, K. Destounis and K. D. Kokkotas, \textit{Charged black holes in de Sitter space:
 superradiant amplification of charged scalar waves and resonant hyperradiation}, Phys.
 Rev. D105, 084052 (2022), \texttt{ arXiv:2204.05335}.





\bibitem{150}

V.  V. Kiselev,  \textit{Quintessence and black holes}, Class. Quant. Grav. 20, 1187 (2003),  
\texttt{gr-qc/0210040}.

\bibitem{151}
 M. Visser,  \textit{The Kiselev black hole is neither perfect fluid, nor is it quintessence}, Class. Quant.
Grav. 37,  045001 (2020), \texttt{arXiv:21908.11058}.
\bibitem{152}
J. P. Morais Gra\c ca, E. Folco Capossoli, H. Boschi-Filho and I. P. Lobo,  \textit{Joule-Thomson
expansion for quantum corrected AdS-Reissner-Nordstrom black holes in a Kiselev spacetime},
Phys. Rev. D 107 (2023) 024045, \texttt{arXiv:2105.04689}.
\bibitem{1520}
A. Belhaj, A. El Balali, W. El Hadri, Y. Hassouni, E. Torrente-Lujan,  \textit{Phase transition and shadow behaviors of quintessential black holes in M-theory/superstring inspired models},   Int.J.Mod.Phys.A 36,  08n09, 2150057 (2021.

\bibitem{153}

A.  Anand, A.  Mishra, P. Channuie, \textit{Stability of Extremal Black Holes and Weak Cosmic Censorship Conjecture in Kiselev Spacetime}, Int.J.Theor.Phys. 64, 166 (2025),   \texttt{arXiv:2411.02427}.
\bibitem{154}
 F.  Ahmed, A.  Al-Badawi, İ. Sakallı, \textit{Quantum Oppenheimer-Snyder Black Hole with Quintessential Dark Energy and a String Clouds: Geodesics, Perturbative Dynamics, and Thermal Properties}, \texttt{arXiv:2508.03202}.


 
\bibitem{17}
R. B. Wang, S. J. Ma, L. You, J. B. Deng and X. R. Hu, \textit{Thermodynamics of Schwarzschild-AdS black hole in non-commutative geometry},  Chin. Phys. C 49 065101 (2025), \texttt{ arXiv:2410.03650}.



  \bibitem{18}
 J. D. Bekenstein,   \textit{Extraction of Energy and Charge from a Black Hole}, Phys. Rev. D, 7,  949
(1973).

  \bibitem{171}
A. Kumar, D.V. Singh, S.G. Ghosh, \textit{ D-dimensional Bardeen-AdS black holes in Einstein-
Gauss-Bonnet theory}, Eur. Phys. J. C79 275(2019).
\bibitem{170}
H. Liu, X. h. Meng, \textit{P--V Criticality In the Extended Phase Space of Charged Accelerating AdS Black Holes}, Mod. Phys. Lett. A, 31 1650199 (2016), \texttt{arXiv:1607.00496}.
\bibitem{1700}
H. Belmahi, M. Jemri, R. Salih, \textit{Stability and Criticality Behaviors of
 Accelerating Charged AdS Black Holes in
 Rainbow Gravity}, \texttt{arXiv:2507.03572}.


\bibitem{21}
 D. Kubiznak, R. B. Mann,\textit{ P--V criticality of charged AdS black holes,} JHEP 033,  1207 (2012).
 
\bibitem{22}
H.~K.~Sudhanshu, D.~V.~Singh, S.~Bekov, K.~Myrzakulov and S.~Upadhyay,
\textit{P-v criticality and Joule-Thomson expansion in corrected thermodynamics of conformally dressed (2+1)D AdS black hole},
Int. J. Mod. Phys. A \textbf{38}  2350165 (2023).
\bibitem{23}
M. Y. Zhang, H. Chen, H. Hassanabad, Z. W. Long, H. Yang, \textit{Critical behavior and Joule-Thomson expansion of charged AdS black holes surrounded by exotic fluid with modified Chaplygin equation of state}, Chinese Physics C 28 6 (2024). 

\bibitem{24}
Y. Meng, J. Pu, Q. Q. Jiang,
\textit{P-V criticality and Joule-Thomson expansion of charged AdS black holes in
the Rastall gravity}, Chinese Physics C  44 6 065105 (2020). 

\bibitem{25}
H. Liu, X. h. Meng, \textit{P--V Criticality In the Extended Phase Space of Charged Accelerating AdS Black Holes}, Mod. Phys. Lett. A, 31 1650199 (2016), \texttt{arXiv:1607.00496}.

\bibitem{27}
 D. Kubiznak, R. B. Mann,\textit{ P--V criticality of charged AdS black holes,} JHEP 033 1207 (2012), \texttt{ arXiv:1205.0559}.
 
\bibitem{28} 
 M. Y. Zhang, H. Chen, H. Hassanabadi, Z. W. Long, H. Yang, \textit{Joule-Thomson expansion of charged dilatonic black holes}, Chin.Phys.C 47  4, 045101 (2023), \texttt{arXiv:2209.00868}.
 
 \bibitem{29}
 M. Yasir, X. Tiecheng, F. Javed, G. Mustafa, \textit{Thermal analysis and Joule-Thomson expansion of black hole exhibiting metric-affine gravity}, Chin.Phys.C 48 1, 015103 (2024), \texttt{arXiv:2305.13709}.
 
\bibitem{30}
   K. Hegde, A. Naveena Kumara, C. L. Ahmed Rizwan, A. K. M., M. S. Ali
 and S. Punacha,\textit{ Thermodynamics,  phase transition and Joule–Thomson
 expansion of 4-D Gauss–Bonnet AdS black hole}, Int. J. Mod. Phys. A 39
 2450080 (2024), \texttt{arXiv:2003.08778}.
 
 \bibitem{31}
Ö. Ökcü, E. Aydıner,  \textit{Joule-Thomson expansion of the charged AdS black hole}, 	Eur. Phys. J. C, 77 (1) 24 (2017),   \texttt{arXiv:1611.06327}.

 \bibitem{32}
M. Chabab, H. El Moumni, S. Iraoui, K. Masmar, S. Zhizeh,   \textit{Joule-Thomson expansion of RN-AdS black holes in f(R) gravity}, Lett. High Energy Phys. 2 (05)
(2018). 
  \bibitem{320}
 K. Masmar,   \textit{Joule–Thomson expansion for a nonlinearly charged Anti-de Sitter black
hole}, Int. J.Geom.Meth.Mod.Phys. 20,  05, 2350080 (2023).
 
   \bibitem{A117}
 H. Belmahi, \textit{Constrained Deflection Angle and Shadows of Rotating
Black Holes in Einstein-Maxwell-scalar Theory},  \texttt{	arXiv:2411.11622}.
 
  \bibitem{A118}
 A. Belhaj, H. Belmahi, M. Benali, \textit{Deflection Light Behaviors by AdS Black Holes,} Gen Rel. Grav. 54, 4 (2022), \texttt{	arXiv:2112.06215}.
 
 
 \bibitem{A119}
 A. Övgün, \textit{Weak field deflection angle by regular black holes with cosmic strings using the Gauss-Bonnet theorem,} Phys. Rev. D 99, 104075 (2019).
 
 
\end{thebibliography}
\end{document}